\documentclass[final,3p,times,authoryear]{elsarticle}

\usepackage{amsmath}
\usepackage{amsfonts}
\usepackage{amssymb}
\usepackage[titletoc,title]{appendix}
\usepackage{booktabs}
\usepackage[mmddyyyy]{datetime}
\usepackage{dutchcal}
\usepackage{enumitem}
\usepackage{float}
\usepackage{graphicx}
\usepackage{hyperref}
\usepackage[utf8]{inputenc}
\usepackage{lineno}
\usepackage{natbib}
\usepackage[percent]{overpic}
\usepackage{pgfplots}
\usepackage{relsize}
\usepackage{subfigure}
\usepackage{siunitx}
\usepackage{tikz}

\newcommand{\ra}[1]{\renewcommand{\arraystretch}{#1}}
\modulolinenumbers[5]
\numberwithin{equation}{section}

\usetikzlibrary{calc}
\usetikzlibrary{plotmarks}
\usetikzlibrary{positioning}
\usetikzlibrary{fit}
\usetikzlibrary{decorations.pathmorphing}
\usetikzlibrary{backgrounds}
\makeatother
\modulolinenumbers[5]
\numberwithin{equation}{section}
\pgfplotsset{compat = newest}
\pgfplotsset{ legend style={font=\tiny} }
\definecolor{bgreen}{rgb}{0.0,0.5,0.0}
\definecolor{bblue}{rgb}{0.0,0.0,0.9}
\definecolor{bgold}{rgb}{0.7,0.5,0.0}
\definecolor{bred}{rgb}{0.9,0.0,0.0}

\begin{document}

\begin{frontmatter}



\title{Self-similar orbit-averaged Fokker-Planck equation for isotropic spherical dense clusters (ii) physical properties and negative heat capacity of pre-collapse core}


\author[address1,address2]{Yuta Ito}
\ead{yito@gradcenter.cuny.edu}

\address[address1]{Department of Physics, CUNY Graduate Center, \tnoteref{footnote1}} 
\address[address2]{Department of Engineering and Physics,
	CUNY College of Staten Island\tnoteref{footnote2}}
\tnotetext[footnote1]{ 365 Fifth Avenue, New York, NY 10016, USA}
\tnotetext[footnote2]{2800 Victory Boulevard, Staten Island, NY 10314, USA}

\begin{abstract}
This is the second paper of a series of our works on the self-similar orbit-averaged Fokker-Planck (OAFP) equation and details physical properties of isotropic pre-collapse solution. The fundamental core collapse process at the late stage of relaxation evolution of spherical star clusters can be described by the self-similar OAFP equation. The accurate spectral solution was found recently in the first paper.  The present work details the thermodynamic aspects of the model based on the stellar DF obtained from the solution. Our calculation shows the following local properties (i) Equation of state is $p=1.0\rho/\chi_\text{esc}$ in the core where $p$ is the pressure, $\rho$ the density and $\chi_\text{esc}$ the scaled escape energy, while it is $p=0.5\rho^{1.1}/\chi_\text{esc}$ at large radii. (ii) If we consider the center of the core is polytropic, the polytropic index is 177. Also, as a global property we construct caloric curves of the model to discuss the heat capacity together with Virial. Special focus is the cause of negative heat capacity of the core; the well-relaxed core can be directly compared to the isothermal sphere with positive heat capacity. Comparing our results to the previous works, we conclude, in the self-similar evolution, the negative heat capacity in the core holds due to collisionless and high-temperature stars that experience a rapid change in mean field potential through stellar- and heat- flows, rather than due to the (quasi-)isolation of the core from surroundings. 

\end{abstract}


\end{frontmatter}

\section{Introduction}

This is the second paper of a series of our works on the self-similar orbit-averaged Fokker-Planck (ss-OAFP) equation for density distribution of stars in isotropic dense star cluster at core collapse phase. The ss-OAFP model is important to understand the late stage of relaxation evolution of the clusters. In the first paper \citep{Ito_2020_1}, we found an accurate Gauss-Chebyshev spectral solution of the equation. Although a self-similar gaseous model \citep{Lynden_Bell_1980} has shown a unique physical feature of star clusters at pre-collapse stage in the sense that it is at a noequilibrium state different from the state of clusters at the early stage of evolution described by the King model \citep{King_1966}, one has yet to reveal the detail physical feature of the pre-collapse state based on kinetic approach, other than a time-dependent OAFP model \citep{Cohn_1980}. Also, the studies on the ss-OAFP model by \citep{Heggie_1988,Takahashi_1993} did not detail the physical feature of the model. Hence, the present work analyzes the model following the method established for equilibrium statistical mechanics of isotropic self-gravitating systems. Section \ref{sec:Intro_rev} explains the basic physical feature of the ss-OAFP model contrasting the model with basic isotropic self-gravitating models. Since the ss-OAFP model essentially describes clusters at a non-equilibrium state rather than equilibrium, Section \ref{sec:Intro_assum} explains the assumptions we made for the model. Also, since the previous works have not detailed the heat capacity of pre-collapse self-similar model, the present work focuses on the negative heat capacity of the ss-OAFP model. Especially our interest is the heat capacity of the core since the core of the ss-OAFP model is supposed to be well-relaxed so the core can be readily compared to heat capacity of the isothermal sphere. While the negative heat capacity of the equilibrium self-gravitating systems and systems of long- and short- range interacting particles have attracted one's concern, many of them focused only on equilibrium states (Section \ref{cause_nCV_comp_reported}). We aim at showing the unique cause of the negative heat capacity of the core of the ss-OAFP model in the present paper.

\subsection{Isotropic self-gravitating system in equilibrium mechanics}\label{sec:Intro_rev}

One may list the isotropic models that could establish equilibrium statistical mechanics of self-gravitating (gaseous) systems based on what temperature one employs for the systems (Table \ref{table:review}).
\begin{table*}\centering
	\ra{1.3}
	\scalebox{0.7}{
	\begin{tabular}{@{}l|ccc@{}}\toprule
		Model&Temperature & entropy    & Ref\\ 
		\midrule
		Isothermal sphere & Thermodynamic  &Boltzmann-Gibbs &  \citep{Antonov_1962,Lynden_Bell_1968}   \\
		Energy-truncated models   & Thermodynamic  &Phenomenological  &  \citep{Katz_1980,Katz_1983}  \\
		(e.g. \emph{Woolly}, \emph{King} and \emph{Wilson}  models)
		 &&&\\
		Stellar Polytrope & Polytropic (constant)  &Tsallis's  &  \citep{Taruya_2002,Chavanis_2002_poly} \\
		ss-OAFP models     &    ?    & ?  &     \\
		\bottomrule
	\end{tabular}
}
	\caption{Typical isotropic self-gravitating systems discussed for study on equilibrium statistical mechanics.}
	\label{table:review}
\end{table*}
The ss-OAFP model would have three different (non-)equilibrium states. In the core the stars could behave like an isothermal sphere due to frequent two-body relaxation events. The inner halo would be a receiver of heat- and stellar (particle-) fluxes due to the escaping stars from the core, resulting in a non-equilibrium state. In the outer halo, self-similar analysis\footnote{More realistic arguments for finite outer halo needs to include the effect of flux at the ridge of cluster, two dimensional (anisotropic) effect, and escapers and escaping stars with high eccentricity under the influence of tidal effects. The discussions for the outer halo are found in \citep[e.g.][]{Michie_1962,Spitzer_1972,Claydon_2019}.} extends the inner halo to infinite radius (without or with very little stellar and heat fluxes) and the outer halo would behave like a collisionless system forming a power-law profile at a state of Tsallis equilibrium \citep[Refer to][for detail of Tsallis's generalized statistical dynamics]{Tsallis_2009}. Although the present work focuses on the core of the ss-OAFP model, to properly discuss the complicated entire structure of the model, one would need to resort to a hybrid entropy of Boltzmann-Gibbs and Tsallis's entropies \citep[e.g.][]{Kang_2011}. 

\subsection{Assumptions made for the ss-OAFP model}\label{sec:Intro_assum}

To discuss thermodynamic aspects of the ss-OAFP model, one needs several assumptions. For example, the system can reasonably exist only in a time-averaged sense due to the discreteness (finiteness of total number of stars) of star clusters. Hence, to consider the system a thermodynamic system, it must be an 'exotic' self-gravitating gas at a stationary non-equilibrium state, composed of $N(>>10^{6}>>1)$ particles whose distribution function (DF) and mean field (m.f.) potential follow the solution of the ss-OAFP model at a certain time of self-similar evolution on relaxation time scale. The meaning of 'exotic' is here that the gas has intrinsically \emph{negative} heat capacity as its normal state (at least in the core of the system) as explained in Sections \ref{sec:Reg_C-curve} and \ref{sec:cause_nCv}. In addition, the ss-OAFP model is not isolated, which implies the model can not reach a QSS in the sense that Virial ratio can not reach unity. Hence one may expect the model may not achieve an equilibrium state but still can be at stationary in the inner- and outer- halos (in the case of self-similar model). For the rest, following the classical discussion \citep{Antonov_1962,Lynden_Bell_1968}, we consider the ss-OAFP model is enclosed by an adiabatic spherical wall of radius $R(\Phi)$, where $\Phi$ is the self-similar form of m.f. potential $\phi(\text{r},t)$, that elastically reflects stars on the inner surface. The present work employs the reference solution $F_\text{o}(E)$ and $\Phi_\text{o}(R)$ as stellar DF and m.f. potential obtained in \citep{Ito_2020_1} for further discussion. The DF $F_\text{o}(E)$ and energy $E$ are the self-similar forms of probability phase-space density DF $f(\epsilon,t)$ for stars and energy $\epsilon(=\textbf{v}^2/2+\phi(\textbf{r},t))$ that is available to a star at phase point $(\textbf{r},\textbf{v})$ orbiting in an isotropic star cluster at time $t$. Also, we assume the time $t$ of the ss-OAFP model is fixed to a certain time $t_\text{c}$. This means the time-dependent variables (e.g. $f_\text{t}, g_\text{t},\epsilon_\text{t},...$) in \citep{Ito_2020_1} turn into the factors that make the self-similar variables (e.g. $F, G, R, ...$) in dimensionless form.

\subsection{The cause of negative heat capacity reported in the previous works}\label{cause_nCV_comp_reported}

Negative heat capacity has been discussed not only for self-gravitating astrophysical objects such as stars, star clusters and black holes \citep{Lynden_Bell_1999} but also stratified gasses \citep{Ingel_2000}, proteins \citep{Prabhu_2006}, granular gasses \citep{Brilliantov_2018} and systems of small $N$ particles such as Lennad-Jones gas \citep{Thirring_2003}, melting metal clusters \citep{Aguado_2011} and hot nuclei \citep{Borderie_2019}. Many of existing works are based on numerical simulations and analytical approaches though, study on the negative heat capacity itself is not just an academic exercise. Laboratory experiments have shown the signature of negative specific capacity of nuclear fragmentation \citep{D_Agostino_1999,D_Agostino_2000,Srivastava_2001,Gobet_2002} and of melting metal clusters \citep{Schmidt_2001}. The existing works commonly focus on microcanonical ensembles and yield that the negative heat capacity occurs in an energy band on which the system undergoes a phase transition such as 'gas'-'liquid' and 'liquid'-'solid'. In the case of star clusters the phase transition occurs for 'gas'-'collapsed-state' in which 'gas' corresponds with a state of self-gravitating gas with positive heat capacity and 'collapsed-state' means a state of core-halo structure with negative heat capacity. For brevity, our interest is negative heat capacity of the core of the ss-OAFP model and it is considered a system of small numbers ($N<<10^6$) of stars.

For discussion of negative heat capacity, one typically concerns with inhomogenous microcanonical ensemble, which may be explained by discussing systems that do \emph{not} undergo negative heat capacity. In the case of a small $N$ system of homogeneous subsystems at  a state of thermodynamic equilibrium e.g. nanoclusters \citep{Michaelian_2007, Lynden_Bell_2008, Michaelian_2015, Michaelian_2017}, negative heat capacity does not occur regardless of kinds of statistical ensembles. Commonly the negative heat capacity has been reported for microcanonical ensemble \citep{Lynden_Bell_1999} though, the small $N$ systems can not achieve ergocity due to particles being generally trapped only in limited part of the whole phase space, meaning less mixing process occurs in the system and the final state (and heat capacity) should depend on the initial condition. On one hand, in the case of inhomogeneous equilibrium systems, negative heat capacity does not occur for canonical ensemble \citep{Thirring_1970, Lynden_Bell_1977} and grandcanonical ensemble \citep{Josephson_1967} since the heat capacities are defined by the square of fluctuation in thermodynamic quantities in the same way as the homogeneous case. These discussions typically suggests negative heat capacity could occur only for (isolated) inhomogeneous system in microcanonical ensemble \citep{Lynden_Bell_2008}. 

Although the negative heat capacity has been found in (nearly) isolated astrophysical systems of particles or stars interacting via Newtonian potential, the long-range nature of pair-wise potential itself is not only the cause of the negativeness. For example, even non-interacting particles can have negative heat capacity under certain background potentials. Based on Virial theorem and toy models, \cite{Einarsson_2004} showed collisionless particles in the background potential profile $\sim r^{a}$ (in three dimension) provides negative heat capacity if $a=-1$ while if $a\neq-1$ it does not. \cite{Thirring_2003,Carignano_2010} showed negative heat capacity can occur to collisionless particles under a sudden change in a potential change from a deep narrow potential well at small radii toward a shallow wide well at larger radii. On one hand, even without a background potential well, particles interacting through short-range pair potential can reveal negative heat capacity. The examples are Lennard-Jones potential for small $N (\sim10)$ \citep{Thirring_2003} and Gaussian potential \citep{Posch_1990,Posch_2005} for $N\sim100$.

We aim at finding the cause of negative specific heat in the core of the ss-OAFP model after detailing the physical feature of the model based on the moments of the reference DF and Virial. The present paper is organized as follows. Section \ref{sec:Thermo_Quan_ss_FP} shows the local and global properties of the ss-OAFP model based on the moments of the reference DF. Section \ref{sec:Reg_C-curve} aims to construct an analogue of caloric curve and show negative heat capacity at constant volume for the model after regularizing the thermodynamic quantities. Section \ref{sec:cause_nCv} explains the cause of the negative heat capacity in the core of the ss-OAFP model by using simple analytical technique together with the Virial of the model and by comparing the core to existing models. Section \ref{sec:Conclusion} is Conclusion.

\section{Thermodynamic quantities of the ss-OAFP model}\label{sec:Thermo_Quan_ss_FP}

The present section extends \citep{Lynden_Bell_1980}'s analyses to the ss-OAFP model, especially focusing on the core. Sections \ref{sec:LTQ} and \ref{sec:GTQ} detail the local- and global- properties of the ss-OAFP model.

\subsection{Local properties of the ss-OAFP model}\label{sec:LTQ}

The stellar DF of the ss-OAFP model may not be even a local Maxwellian DF, hence we need to find the moments of the DF (i.e. local thermodynamic- or hydrodynamic- quantities)  to capture the physical features of the model. In addition to density $\rho(r,t)$ and its self-similar form $D(\Phi)$ (calculated for the ss-OAFP equation in \citep{Ito_2020_1}, one may define the velocity dispersion, pressure and heat- and stellar- fluxes in terms of moments of DF
\begin{subequations}
	\begin{align}
	&\varv^{2}(r,t_\text{c})=\frac{1}{\rho(r,t_\text{c})}\int^{0}_{\phi(r,t_\text{c})}(2\epsilon-2\phi(r,t_\text{c}))^{3/2}f(\epsilon,t_\text{c})\,\,\text{d}\epsilon,\\
	&p(r,t_\text{c})=\frac{\varv^{2}(r,t_\text{c})\rho(r,t_\text{c})}{3},\\
	&\mathcal{f}^\text{h}(r,t_\text{c})=-\mathcal{k}\frac{\partial \varv^{2}(r,t_\text{c})}{\partial{r}},\\
	&\mathcal{f}^\text{p}(r,t_\text{c})=-\mathcal{D}\frac{\partial \rho(r,t_\text{c})}{\partial{r}},
	\end{align}
\end{subequations}
where $\mathcal{k}$ is the thermal conductivity and $\mathcal{D}$ the diffusivity. Since the value of $\mathcal{k}$ (and $\mathcal{D}$) depends on the definition of relaxation time concerned \citep{Lynden_Bell_1980,Louis_1991} we regularize the heat- and stellar fluxes by the constants in the coefficients in self-similar analysis (that is, by $3GmC_\text{k} \ln[N]$ for the conductivity and $\sqrt{2 C_\text{D}\ln[N]/3}$ for the diffusivity\footnote{For the conductivity the expression of the constants obeyed that of \citep{Lynden_Bell_1980}'s work. In a similar way, the diffusivity can be calculated as follows
	\begin{align}
	\mathcal{D}=\frac{1}{3}l\sqrt{\varv^{2}}\approx\frac{1}{3k_{J}^{2}}\frac{C_\text{D}}{T_{R}}=\frac{1}{3}\frac{\varv^{2}(r,t_\text{c})}{4\pi G \rho(\phi(r,t_\text{c}))}\frac{8\pi Gm C_\text{D}\rho(\phi,t_\text{c})\ln[N]}{[\varv^{2}(r,t_\text{c})]^{3/2}}=\frac{2C_\text{D}\ln[N]}{3\sqrt{\varv^{2}(r,t_\text{c})}},
	\end{align}
	where $l$ is the mean free path of stars, $k_\text{J}$ is the inverse of Jeans length and $T_{R}$ is the relaxation time from \citep{Spitzer_1988}.}, where $m$ is stellar mass, $G$ is the gravitational constant and $C_\text{k}$ and $C_\text{D}$ are dimensionless constants of approximately unity.). Employing the self-similar variables in \citep{Ito_2020_1} and introducing new dimensionless variables
\begin{subequations}
	\begin{align}
	&\varv^{2}(r,t_\text{c})=V^{2}_\text{(dis)}(R)\varv^{2}_\text{t}(t)\equiv 2V_\text{(dis)}^{2}(R)\epsilon_{t}(t),\\
	&p(r,t_\text{c})=P(R)p_\text{t}(t),\\
	&\frac{\mathcal{f}^\text{h}(r,t_\text{c})}{3GmC_\text{k} \ln[N]}\equiv\mathcal{F}^\text{h}(R)\mathcal{f}^\text{h}_\text{t}(t_\text{c}),\\
	&\frac{\mathcal{f}^\text{p}(r,t_\text{c})}{2C_\text{D} \ln[N]/3}\equiv\mathcal{F}^\text{p}(R)\mathcal{f}^\text{p}_\text{t}(t_\text{c}),
	\end{align}
\end{subequations}
the local quantities in dimensionless form read
\begin{subequations}
	\begin{align}
	&V_\text{(dis)}^{2}(R)=2\frac{U_{3/2}(\Phi)}{U_{1/2}(\Phi)},\\
	&P(R)=\frac{2}{3}U_{3/2}(\Phi),\\
	&\mathcal{F}^\text{h}(R)=-\frac{U_{1/2}(\Phi)}{S(\Phi)\sqrt{V_\text{(dis)}^{2}(\Phi)}}\left(\frac{U_{3/2}(\Phi)U_{-1/2}(\Phi)}{[U_{1/2}(\Phi)]^{2}}-3\right),\\
	&\mathcal{F}^\text{p}(R)=-\frac{1}{\sqrt{V_\text{(dis)}^{2}(\Phi)}}\frac{U_{-1/2}(\Phi)}{S(\Phi)},
	\end{align}
\end{subequations}
where $S(\Phi)=-\frac{\text{d}R}{\text{d}\Phi}(x)$ and $U_{n}(\Phi)$ is
\begin{align}
U_{n}(\Phi)=\int^{0}_{\Phi}F(E)(E-\Phi)^{n}\,\text{d}E,
\end{align}
where $n$ is a real number; for example, if $n=1/2$  then $U_{n}(\Phi)=D(\Phi)$. 

\begin{figure}[H]
	\centering
	\begin{tikzpicture}
	\begin{loglogaxis}[width=7cm,height=7cm,
	grid=major,xlabel=\Large{$R$},xmin=1e-1,xmax=1e5,ymin=1e-2,ymax=1.2e0,legend pos=south west]
	\addplot [color = red ,mark=no,thick,densely dashed ] table[x index=0, y index=1]{R_Phi_Vdis_HeatF_DensF_InvTr.txt};
	\addlegendentry{\large{$\mid\Phi\mid$}} 
	\addplot [color = blue ,mark=no,thick,solid ] table[x index=0, y index=1]{R_LocT_KinT.txt};
	\addlegendentry{\large{$V\text{(dis)}^{2}$}}
	\node[above,black] at (4e4,0.7) {\Large{\textbf{(a)} }};
	\end{loglogaxis}
	\end{tikzpicture}\hspace{0.3cm}
	\begin{tikzpicture}
	\begin{loglogaxis}[width=7cm,height=7cm,
	grid=major,xlabel=\Large{$R$},xmin=1e-1,xmax=1e8,ymin=1e-24,ymax=1.1e1,legend pos=south west]
	\addplot [color = red ,mark=no,thick,solid ] table[x index=0, y index=3]{R_Phi_Vdis_HeatF_DensF_InvTr.txt};
	\addlegendentry{\Large{$\mathcal{F}^\text{h}$}} 
	\addplot [color = orange ,mark=no,thick,densely dashed ] table[x index=0, y index=4]{R_Phi_Vdis_HeatF_DensF_InvTr.txt};
	\addlegendentry{\Large{$\mathcal{F}^\text{p}$}} 
	\addplot [color = blue ,mark=no,thick,densely dotted] table[x index=0, y index=5]{R_Phi_Vdis_HeatF_DensF_InvTr.txt};
	\addlegendentry{\Large{$1/\tilde{T}_\text{R}$}} 
	\node[above,black] at (4e7,1e-2) {\Large{\textbf{(b)}}};
	\end{loglogaxis}
	\end{tikzpicture}
	\caption{\textbf{(a)} Dimensionless m.f. potential $\Phi$ and velocity dispersion $V_\text{(dis)}^{2}$ and \textbf{(b)} dimensionless heat flux $\mathcal{F}^\text{h}$ and stellar flux $\mathcal{F}^\text{p}$. The latter also depicts the inverse of regularized local relaxation time $\tilde{T}_\text{R}$ $\left(=D(\phi,t_\text{c})/[\varv^{2}(r,t_\text{c})]^{3/2}\right)$ for comparison.}
	\label{Fig.R_Phi_Vdis_HeatF_DensF_Tr}
\end{figure}
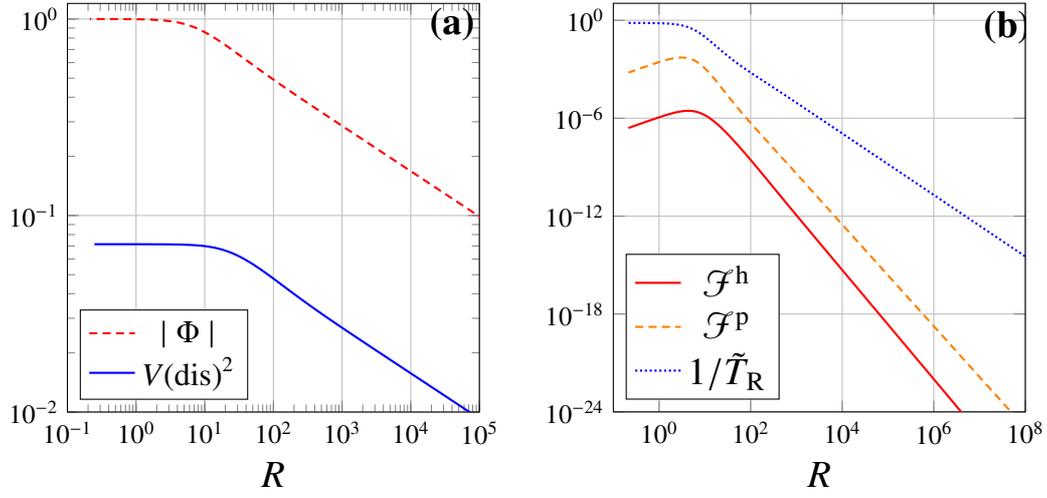

The ss-OAFP model has characteristics similar to those of the self-similar conductive gaseous model reported in \citep{Lynden_Bell_1980}. Figure \ref{Fig.R_Phi_Vdis_HeatF_DensF_Tr}\textbf{(a)} depicts the velocity dispersion $V_\text{(dis)}^{2}$ and m.f. potential profile $\Phi$.  The constancy of $V_\text{(dis)}^{2}$ further extends in radius compared to that of $\Phi$. Figure \ref{Fig.R_Phi_Vdis_HeatF_DensF_Tr}\textbf{(b)} shows  the heat flux $\mathcal{F}^\text{h}$ and stellar flux $\mathcal{F}^\text{p}$. The flux $\mathcal{F}^\text{h}$ reaches its maximum at $R=4.31$. This radius is relatively close to $R=8.99$ at which the gravothermal instability occurs for the isothermal sphere in canonical ensemble \citep{Lynden_Bell_1968}. The flux $\mathcal{F}^\text{p}$ reaches its maximum at slightly smaller radius, $R=3.11$. The fluxes $\mathcal{F}^\text{h}$ and $\mathcal{F}^\text{p}$ rapidly decrease as the local relaxation time increases with radius $R$, while $\mathcal{F}^\text{p}$ decays slightly slowly compared to $\mathcal{F}^\text{h}$. The location of the maximum stellar flux can be explained by the escape speed of stars yet, which is discussed in Section \ref{sec:GTQ} (since the speed is a global property of the cluster.) The tuning points for fluxes seem slightly different from the graphed maximum value in \citep{Lynden_Bell_1980} but the qualitative nature of our velocity dispersion, m.f. potential and fluxes little differ from the gaseous model.

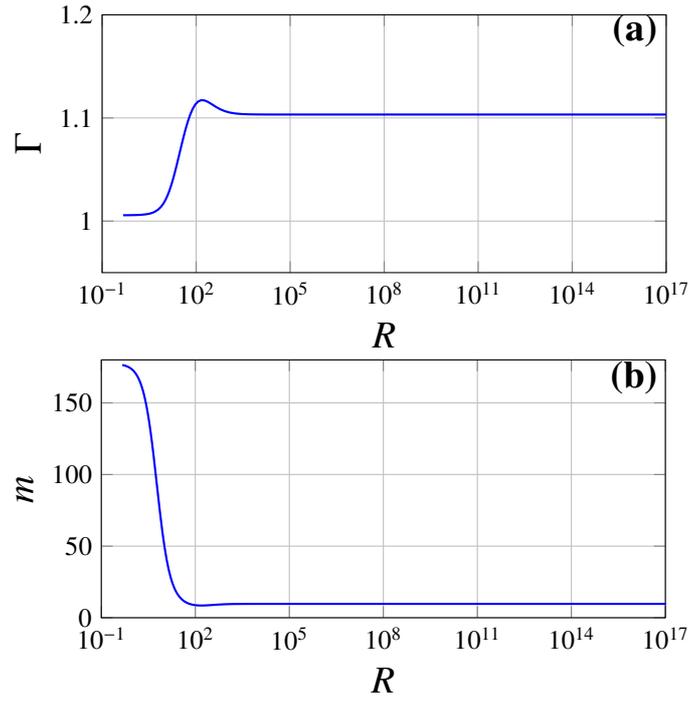
\begin{figure}[H]
	\centering
	\begin{tikzpicture}
	\begin{semilogxaxis}[width=9cm,height=5cm,
	grid=major,xlabel=\Large{$R$},ylabel=\Large{$\Gamma$},xmin=1e-1,xmax=1e17,ymin=9.5e-1,ymax=1.2e0,legend pos=south east]
	\addplot [color = blue ,mark=no,thick,solid] table[x index=0, y index=1]{R_AdiaIndex.txt};
	\node[above,black] at (1e16,1.16) {\Large{\textbf{(a)} }};
	\end{semilogxaxis}
	\end{tikzpicture}
	\begin{tikzpicture}
	\begin{semilogxaxis}[width=9cm,height=5cm,
	grid=major,xlabel=\Large{$R$},ylabel=\Large{$m$},xmin=1e-1,xmax=1e17,ymin=0,ymax=180,legend pos=south east]
	\addplot [color = blue ,mark=no,thick,solid] table[x index=0, y index=1]{R_m_ss.txt};
	\node[above,black] at (1e16,150) {\Large{\textbf{(b)} }};
	\end{semilogxaxis}
	\end{tikzpicture}
	\caption{ \textbf{(a)} Adiabatic index $\Gamma$ and \textbf{(b)} polytropic index $m$ of the ss-OAFP model.}
	\label{Fig.R_AdiaIndex_m_ss}
\end{figure}

\begin{figure}[H]
	\centering
	\begin{tikzpicture}
	\begin{semilogxaxis}[width=9cm,height=5cm,
	grid=major,xlabel=\Large{$R$},ylabel=\Large{$P\chi_\text{esc}/D^{\Gamma}$},xmin=1e-1,xmax=1e18,ymin=0.1,ymax=1.2e0,legend pos=north east]
	\addplot [color = red ,mark=no,thick,solid ] table[x index=0, y index=1]{R_LocEqnState.txt};
	\end{semilogxaxis}
	\end{tikzpicture}\caption{Local equation of state with constant thermodynamic temperature $\chi_\text{esc}$.}
	\label{Fig.R_LocEqnState}
\end{figure}
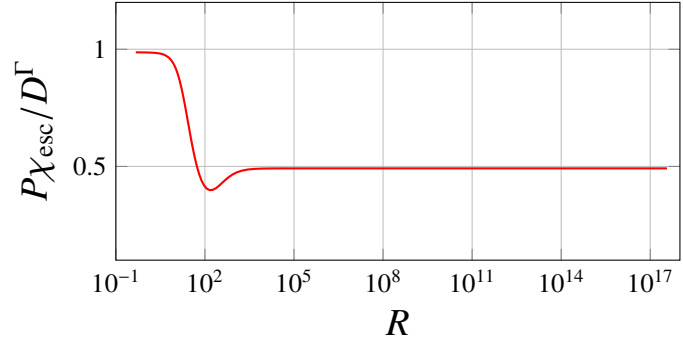

The present work further details the self-similar model based on adiabatic index and equation of local state that have not been discussed in detail for self-similar models. Following \citep{Cohn_1980}'s analyses on time-dependent OAFP model, we calculated the polytropic index $m\equiv\beta+3/2$ of ss-OAFP model.  Figure  \ref{Fig.R_AdiaIndex_m_ss} depicts the adiabatic index $\Gamma (\equiv1+1/m)$ and polytropic index $m$ against dimensionless radius $R$.  We calculated $\Gamma$ taking the logarithmic derivative $\,\text{d}\ln[P(R)]/\,\text{d}\ln[D(R)]$. The result shows a distinctive nature in the structures of the core, inner-halo and outer-halo. In the outer halo, $m$ asymptotically reaches $9.67837115... (\Gamma\approx1.10)$  for large $R$. \footnote{The value of $m$ holds the relative error $6.3\times10^{-10}\%$  from the expected value $\beta_\text{o}+3/2$.} For small $R$ or in the core and inner halo, one can find two two features. (i) The index $m$ reaches its minimum value $8.52$ $(\Gamma=1.117)$ at $R=157$ and (ii) the index $m$ increases with small decreasing $R$ and reaches $177$ $ (\Gamma=1.00564)$ at the center of the core. Higher $m$ (or $\Gamma$ close to unity) implies the core behaves like an isothermal sphere. Given $\Gamma$, one can find the asymptotic form for the equation of local state (Figure \ref{Fig.R_LocEqnState}). As expected, the core is approximately an ideal gas with the thermodynamic temperature $1/\chi_\text{esc}$ while the outer halo behaves like a polytrope of $m=\beta_\text{o}+3/2$ and has a temperature approximately half of $1/\chi_\text{esc}$ given that the polytropic constant $K$ (in $p(r,t_\text{c})=K\rho(r,t_\text{c})^{\Gamma}$) is considered the constant temperature of the polytrope;
\begin{subequations}
	\begin{align}
	&P= \frac{1.0}{\chi_\text{esc}}D, \hspace{1.3cm} (R\approx 0)\\
	&P= \frac{0.5}{\chi_\text{esc}}D^{1.1}, \hspace{1cm} (R\to \infty)
	\end{align}
\end{subequations} 
where the inverse temperature of the ss-OAFP model is $\chi_\text{esc}$.

\subsection{Global properties of the ss-OAFP model}\label{sec:GTQ}

Global properties of the ss-OAFP model can provide understanding of the macroscopic structure as comparison to local properties discussed in Section \ref{sec:LTQ}. Yet, it is hard to conceptualize some thermodynamic quantities for non-equilibrium state, hence the present work avoids exactly defining entropy and the thermodynamic zeroth law. The total- mass, kinetic energy and potential energy of stars that are confined by an adiabatic wall at radius $R_\text{M}$ read

\begin{subequations}
	\begin{align}
	&M(R_\text{M},t_\text{c})=\iint m f(\epsilon,t_\text{c})\,\text{d}^{3}\varv\,\,\text{d}^{3}r,\\
	&\text{KE}(R_\text{M},t_\text{c})=\iint m\frac{\varv^{2}}{2}f(\epsilon,t_\text{c})\,\text{d}^{3}\varv\,\,\text{d}^{3}r,\\
	&\text{PE}(R_\text{M},t_\text{c})=\iint m\phi(r,t_\text{c})f(\epsilon,t_\text{c})\,\text{d}^{3}\varv\,\,\text{d}^{3}r,\\
	&E_\text{tot}(R_\text{M},t_\text{c})=\text{KE}+\text{PE}.
	\end{align}
\end{subequations}
The dimensionless forms are 
\begin{subequations}
	\begin{align}
	&M(\Phi_\text{M})\equiv\frac{M(r,t_\text{c})}{M_\text{t}(t_\text{c})}=-\int^{\Phi_\text{M}}_{-1}DR^{2}S\,\text{d}\Phi,\\
	&\text{KE}(\Phi_\text{M})\equiv\frac{\text{KE}(r,t_\text{c})}{\text{PE}_\text{t}(t_\text{c})}=-\int^{\Phi_\text{M}}_{-1}U_{3/2}R^{2}S\,\text{d}\Phi,\\
	&\text{PE}(\Phi_\text{M})\equiv\frac{\text{PE}(r,t_\text{c})}{\text{PE}_\text{t}(t_\text{c})}=\frac{1}{2}\int^{\Phi_\text{M}}_{-1}\Phi DR^{2}S\,\text{d}\Phi,\\
	&E_\text{tot}(\Phi_\text{M})\equiv\frac{E_\text{tot}(r,t_\text{c})}{PE_\text{t}(t_\text{c})},\label{Eq.Etot_ss}
	\end{align}
\end{subequations}
where $\Phi_{M}$ is the potential at the wall radius $R_{M}$ and the corresponding variables at time $t_c$ have the following relation
\begin{subequations}
	\begin{align}
	&M_\text{t}(t_\text{c})= (4\pi)^{2}\sqrt{2} m f_\text{t}(t_\text{c})[E_\text{t}(t_\text{c})]^{3/2}[r_\text{t}(t_\text{c})]^{3},\\
	&\text{PE}_\text{t}(t_\text{c})=\frac{\text{KE}_\text{t}(t_\text{c})}{2}=\frac{1}{2}M_\text{t}(t)E_\text{t}(t_\text{c}).
	\end{align}
\end{subequations}

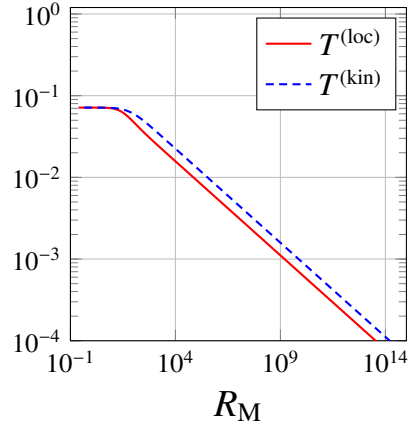
\begin{figure}[H]
	\centering
	\begin{tikzpicture}
	\begin{loglogaxis}[width=6cm,height=6cm,
	grid=major,xlabel=\Large{$R_\text{M}$},xmin=1e-1,xmax=1e15,ymin=1e-4,ymax=1.2e0,legend pos=north east]
	\addplot [color = red ,mark=no,thick,solid ] table[x index=0, y index=1]{R_LocT_KinT.txt};
	\addlegendentry{\large{$ T^\text{(loc)}$}}
	\addplot [color = blue ,mark=no,thick,densely dashed ] table[x index=0, y index=2]{R_LocT_KinT.txt};
	\addlegendentry{\large{$ T^\text{(kin)}$}}
	\end{loglogaxis}
	\end{tikzpicture}\caption{Local temperature and kinetic temperature.}
	\label{Fig.R_LocT_KinT}
\end{figure}

Since the thermodynamic temperature $\chi_\text{esc}$ can not be defined properly as a global quantity for the ss-OAFP model, one may heuristically introduce the kinetic temperature $T^\text{(kin)}$ and local temperature $T^\text{(loc)}$
\begin{subequations}
	\begin{align}
	&T^\text{(kin)}(\Phi_\text{M})\equiv\frac{2\text{PE}}{3M},\\
	&T^\text{(loc)}(R_\text{M})\equiv\frac{V_\text{(dis)}^{2}}{3}.
	\end{align}
\end{subequations}
Figure \ref{Fig.R_LocT_KinT} depicts the local- and kinetic- temperatures. At the center of the system the temperatures hold approximately a constant profile.\footnote{Of course exactly speaking, the temperatures gradually decreases with radius and the maximum temperatures are lowered by 1 $\%$ approximately at $R=2.1$ for local temperature and $R=2.7 $ for kinetic one.} Hence, if focusing on the behavior in the core, one may approximately define the inverse temperature of Maxwellian DF in standard context of statistical mechanics as follows 
\begin{align}
\beta^\text{(con)}\equiv\frac{\chi_\text{esc}}{\epsilon_{t}(t_\text{c})},
\end{align}
This provides an approximate relationship at the center of the system
\begin{subequations}
	\begin{align}
	&T^\text{(loc)}R_\text{M}r_\text{t}(t_\text{c})\approx T^\text{(kin)}R_\text{M}r_\text{t}(t_\text{c})\approx \frac{1}{k_\text{B}\beta^\text{(con)}},\\
	&T^\text{(loc)}(\Phi_{M})\approx T^\text{(kin)}(\Phi_{M})\approx \chi_\text{esc},
	\end{align}
\end{subequations}
where $k_\text{B}$ is the Boltzmann constant. 

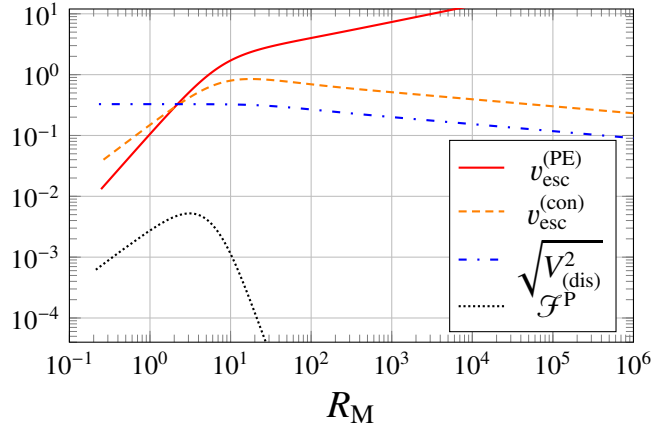
\begin{figure}[H]
	\centering
	\begin{tikzpicture}
	\begin{loglogaxis}[width=9cm,height=6cm,
	grid=major,xlabel=\Large{$R_\text{M}$},xmin=1e-1,xmax=1e6,ymin=4e-5,ymax=1.2e1,legend pos=south east]
	\addplot [color = red ,mark=no,thick,solid ] table[x index=0, y index=1]{R_VescTOT_VescCON.txt};
	\addlegendentry{\large{$ \varv_\text{esc}^\text{(PE)}$}}
	\addplot [color = orange ,mark=no,thick,densely dashed ] table[x index=0, y index=2]{R_VescTOT_VescCON.txt};
	\addlegendentry{\large{$ \varv_\text{esc}^\text{(con)}$}}
	\addplot [color = blue ,mark=no,thick,loosely dashdotted] table[x index=0, y index=2]{R_Phi_Vdis_HeatF_DensF_InvTr.txt};
	\addlegendentry{\large{$ \sqrt{V_\text{(dis)}^{2}}$}}
	\addplot [color = black ,mark=no,thick, densely dotted  ] table[x index=0, y index=4]{R_Phi_Vdis_HeatF_DensF_InvTr.txt};
	\addlegendentry{\large{$ \mathcal{F}^\text{P}$}}
	\end{loglogaxis}
	\end{tikzpicture}\caption{Escape speeds of stars. As comparison, the stellar flux $ \mathcal{F}^\text{P}$ and velocity dispersion $ \sqrt{V_\text{(dis)}^{2}}$ are depicted.}
	\label{Fig.R_VescTOT_VescCON}
\end{figure}

One also can calculate the escape speed from the cluster to explain the maximum and rapid decay of the stellar flux $ \mathcal{F}^\text{P}$ in Figure \ref{Fig.R_Phi_Vdis_HeatF_DensF_Tr}. First, we introduce the following escape speeds in dimensionless form 
\begin{subequations}
	\begin{align}
	\varv_\text{esc}^\text{(PE)}=\sqrt{-4\text{PE}},\\
	\varv_\text{esc}^\text{(con)}=\sqrt{4\frac{M}{R_\text{M}}},
	\end{align}
\end{subequations}
where $\varv_\text{esc}^\text{(con)}$ is introduced based on Virial theorem considering the core is the isothermal sphere isolated from the halo. Figure \ref{Fig.R_VescTOT_VescCON} compares the escape speeds to stellar flux $ \mathcal{F}^\text{P}$. In the core,  velocity dispersion $\sqrt{V_\text{(dis)}^{2}}$  is greater than the escape speeds. Yet, at the center of the core there is less mass flux due to the flattening in density. On one hand, as $R_\text{M}$ increases,  decreasing density (or the transition to inner halo from the core) causes increasing stellar flux. Beyond  $R_M=2.2\sim2.3$ at which $\sqrt{V_\text{(dis)}^{2}}$ is order of the escape speeds, stars can hardly escape due to the self-gravity of the cluster and less encounters occur hence stellar flux decays rapidly with $R_\text{M}$.  

\section{Regularized thermodynamic quantities and negative heat capacity}\label{sec:Reg_C-curve}

The present section focuses on the core of the ss-OAFP model to discuss the negative heat capacity. Section \ref{sec:reg_thermo} regularizes the total energy and temperatures of the model and Section \ref{sec:C-curve} depicts caloric curves to characterize the negative heat capacity.

\subsection{Regularized thermodynamic quantities}\label{sec:reg_thermo}

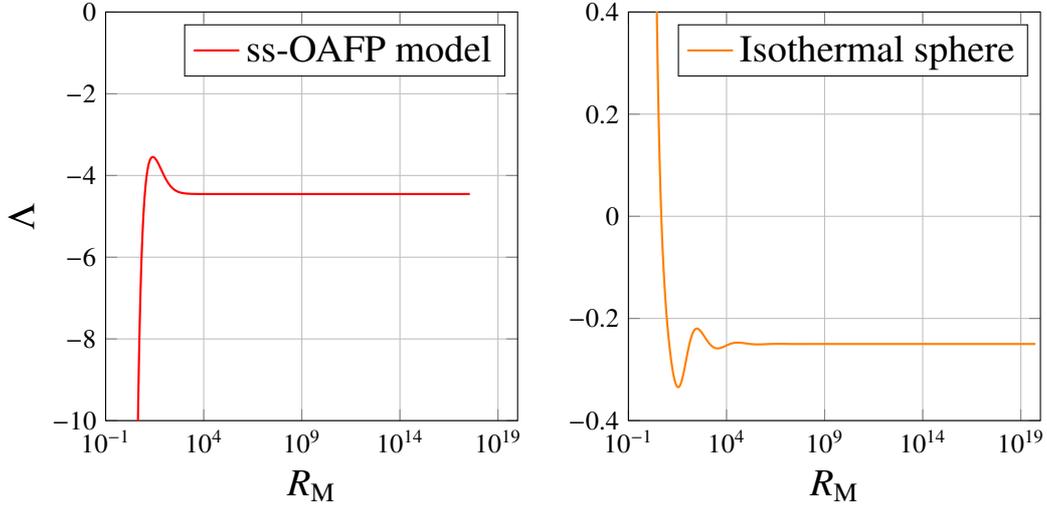
\begin{figure}[H]
	\centering
	\begin{tikzpicture}
	\begin{semilogxaxis}[width=7cm,height=7cm,
	grid=major,xlabel=\Large{$R_\text{M}$},ylabel=\Large{$\Lambda$},xmin=1e-1,xmax=1e20,ymin=-10,ymax=0,legend pos=north east]
	\addplot [color = red ,mark=no,thick,solid ] table[x index=0, y index=1]{R_Lambda.txt};
	\addlegendentry{\Large{ss-OAFP model}}
	\end{semilogxaxis}
	\end{tikzpicture}\hspace{0.3cm}
	\begin{tikzpicture}
	\begin{semilogxaxis}[width=7cm,height=7cm,
	grid=major,xlabel=\Large{$R_\text{M}$},xmin=1e-1,xmax=1e20,ymin=-0.4,ymax=0.4,legend pos=north east]
	\addplot [color = orange ,mark=no,thick,solid ] table[x index=0, y index=1]{Iso_r_Lambda_eta.txt};
	\addlegendentry{\Large{Isothermal sphere}}
	\end{semilogxaxis}
	\end{tikzpicture}
	\caption{Dimensionless normalized total energies for the ss-OAFP model and isothermal sphere. As $R_\text{M}\to0$ the energy monotonically decreases for the former and increases for the latter. The curve for the isothermal sphere was obtained using the numerical code in \citep{Ito_2018}.}
	\label{Fig.R_Lambda_Eta}
\end{figure}

\begin{figure}[H]
	\centering
	\begin{tikzpicture}
	\begin{loglogaxis}[width=9cm,height=4cm,
	grid=major,ylabel=\Large{$1/\eta^\text{(kin)}$},xmin=1e-1,xmax=1e20,ymin=1e-1,ymax=1e3,legend pos=north east]
	\addplot [color = red ,mark=no,thick,solid ] table[x index=0, y index=1]{R_nTkin_nTloc_nTcon.txt};
	\addlegendentry{\Large{ss-OAFP model}}
	\end{loglogaxis}
	\end{tikzpicture}
	
	\vspace{0.3cm}
	
	\begin{tikzpicture}
	\begin{loglogaxis}[width=9cm,height=4cm,
	grid=major,ylabel=\Large{$1/\eta^\text{(loc)}$},xmin=1e-1,xmax=1e20,ymin=1e-1,ymax=1e3,legend pos=north east]
	\addplot [color = red ,mark=no,thick,solid ] table[x index=0, y index=2]{R_nTkin_nTloc_nTcon.txt};
	\addlegendentry{\Large{ss-OAFP model}}
	\end{loglogaxis}
	\end{tikzpicture}
	
	\vspace{0.3cm}
	
	\begin{tikzpicture}
	\begin{loglogaxis}[width=9cm,height=4cm,
	grid=major,ylabel=\Large{$1/\eta^\text{(con)}$},xmin=1e-1,xmax=1e20,ymin=1e-1,ymax=1e4,legend pos=south east]
	\addplot [color = red ,mark=no,thick,solid ] table[x index=0, y index=3]{R_nTkin_nTloc_nTcon.txt};
	\addlegendentry{\Large{ss-OAFP model}}
	\end{loglogaxis}
	\end{tikzpicture}
	
	\vspace{0.3cm}
	
	\begin{tikzpicture}
	\begin{loglogaxis}[width=9cm,height=4cm,
	grid=major,xlabel=\Large{$R_\text{M}$},ylabel=\Large{$1/\eta$},xmin=1e-1,xmax=1e20,ymin=1e-1,ymax=1e3,legend pos=north east]
	\addplot [color = orange ,mark=no,thick,solid ] table[x index=0, y index=3]{Iso_r_Lambda_eta_Inveta.txt};
	\addlegendentry{\Large{Isothermal sphere}}
	\end{loglogaxis}
	\end{tikzpicture}
	\caption{Dimensionless temperatures of the ss-OAFP model and temperature $1/\eta$ of the isothermal sphere. The curve for the latter was obtained using the numerical code in \citep{Ito_2018}}
	\label{Fig.R_nTkin_nTloc_nTcon}
\end{figure}
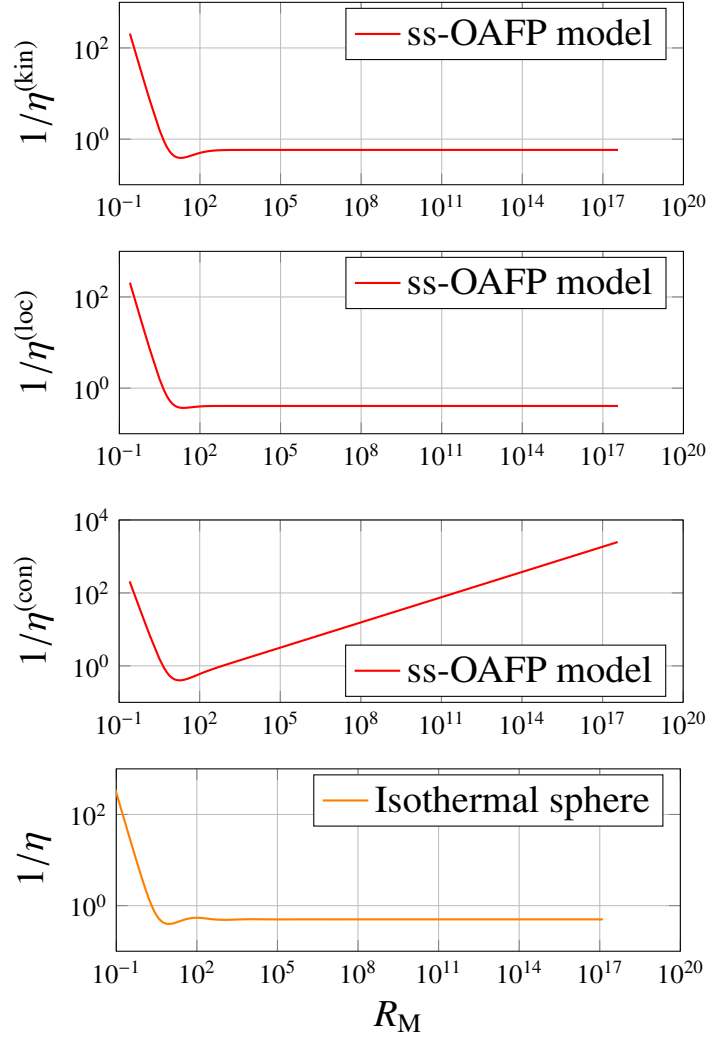

In the present section, we regularize the total energy and temperatures to include into our discussion the effect of finite size of the system enclosed by the wall of radius $R_M$. One would like to consider even for inhomogeneous thermodynamic systems of log-range interacting stars (particles) that there exists a thermodynamic limit in the same way as the standard limit $N/V\to$constant if $N\to\infty$ and $V\to\infty$ for homogeneous systems of particles undergoing short-range interaction. Also, since the mass, energy and size of the ss-OAFP model are infinite, one must resort to a proper thermodynamic limit to regularize thermodynamic quantities at the wall radius $R_\text{M}$. For total energy, the following normalization has been employed for the isothermal sphere based on Boltzmann entropy \citep{Antonov_1962,Lynden_Bell_1968} and stellar polytropes based on Tsallis entropy \citep{Taruya_2002,Chavanis_2002_poly}
\begin{align}
\Lambda=\frac{rE_\text{tot}(r,t_\text{c})}{GM^{2}(r,t_\text{c})}=\frac{R_\text{M}E_\text{tot}(\Phi_{M})}{\left[M(\Phi_{M})\right]^{2}}.\label{Eq.norm_E}
\end{align}
This regularization originates from the feature of Poisson equation (Appendix \ref{App.norm_Etot}). Figure \ref{Fig.R_Lambda_Eta} shows the regularized total energy $\Lambda$ of the ss-OAFP model, compared to that of the isothermal sphere. The energy of the ss-OAFP model is negative at all radii, meaning that the system is dominated by the potential energy. Statistical dynamicists term this state (or turning point) a 'collapsed phase' that occurs to the isothermal sphere (and polytropes of $m>5$) enclosed by an adiabatic wall of a large radius. The energy $\lambda$ reaches order of unity under the following relation with the limit $N\to\infty$ \footnote{Sometimes the relation and limit are called the \emph{dilute} limit\citep[e.g.][]{de_Vega_2002,Destri_2007} that corresponds to one of Boltzmann-Knudsen limits in standard gaseous kinetic theory.}
\begin{subequations}
	\begin{align}
	&R_\text{M}\propto N,\\
	&M(\Phi_{M})\propto N,\\
	&E_\text{tot}(\Phi_{M})\sim \text{KE}(\Phi_{M})\sim \text{PE}(\Phi_{M})\propto N,\\
	&1/\beta^\text{(con)}\propto 1,
	\end{align}
\end{subequations}
This limit determines only the relations among the magnitudes of quantities. The important point is that there exists some reasonable dimensionless parameter as $R_M \to\infty$.\footnote{For example, the density (cluster) expansion in thermodynamic limit for systems of short-range interacting particles does not mean a density $n$ itself is the expansion parameter, rather the corresponding (dimensionless) occupation number $\frac{4}{3}na^3$, where $a$ is the characteristic size of particles, is the actual parameter.} In a similar way, one can convert the reciprocal of the kinetic-, local- and thermodynamic- temperatures to dimensionless forms
\begin{subequations}
	\begin{align}
	&\eta^\text{(kin)}=\frac{GM(R_\text{M}r_\text{t}(t_\text{c}),t_\text{c})}{k_\text{B}T^\text{(kin)}(R_\text{M}r_\text{t}(t_\text{c}))}=\frac{M}{T^\text{(kin)}R_\text{M}},\\
	&\eta^\text{(loc)}=\frac{GM(R_\text{M}r_\text{t}(t_\text{c}),t_\text{c})}{k_\text{B}T^\text{(loc)}(R_\text{M}r_\text{t}(t_\text{c}))}=\frac{M}{T^\text{(loc)}R_\text{M}},\\
	&\eta^\text{(con)}=\frac{\beta^\text{(con)} R_\text{M}r_\text{t}(t_\text{c})}{GM(R_\text{M}r_\text{t}(t_\text{c}))^2}=-\frac{S(\Phi_\text{M})}{R_\text{M}}.
	\end{align} \label{Eq.norm_T}
\end{subequations}
Figure \ref{Fig.R_nTkin_nTloc_nTcon} depicts the dimensionless inverse temperatures $\eta^\text{(kin)}$, $\eta^\text{(loc)}$ and $\eta^\text{(con)}$ of the ss-OAFP model, compared to the corresponding temperature $\eta$ of the isothermal sphere. All the temperatures monotonically decrease as $R_\text{M}$ increases until they reach their minimum values. Only $1/\eta^\text{(con)}$ diverges at large $R_\text{M}$, which implies that the inverse temperature $\beta^\text{(con)}$ is not correctly regularized. Since the ss-OAFP model behaves like a polytropic sphere of $m=\beta_\text{o}+3/2$ at large $R_\text{M}$, one must regularize $\beta^\text{(con)}$ using the regularization employed in \citep{Taruya_2002,Chavanis_2002_poly} to hold constant (polytropic) temperature.

\subsection{Caloric curve}\label{sec:C-curve}

A caloric curve provides the characteristics of heat capacity of the ss-OAFP model.  The dimensionless energy, equation \eqref{Eq.norm_E}, and temperatures, equation \eqref{Eq.norm_T}, construct caloric curves at constant volume (Figure \ref{Fog.R_InvLamb_INVnTkin_INVnTloc_INVnTcon}). Based on Legendre transformation, the graphs relate to the heat capacity $C_\text{V}$. For example, the kinetic temperature relates to $C_\text{V}$ as follows 
\begin{align}
\left(\frac{\partial \frac{1}{\Lambda}}{\partial \eta^\text{(kin)}}\right)_{R,M}=\left(\frac{\partial \frac{1}{\Lambda}}{\partial \eta^\text{(kin)}}\right)_{V,N_\text{tot}}=\frac{C_\text{V}\left[T^\text{(kin)}\right]^{2}}{ME_\text{tot}^{2}}.
\end{align}
Hence, the slope of the tangent to the caloric curve shows the sign of the heat capacity. The advantage of the caloric curve approach is to provide a simple topological understanding for the (linear) stability of systems in statistical (e.g. micro-canonical- and canonical-) ensembles without solving the corresponding eigenvalue problem \citep{Katz_1978,Katz_1979,Katz_1980}.  Since the ss-OAFP model does not achieve an equilibrium system, the rest of the present section discusses the heat capacity of the ss-OAFP model and its singularity to analogically understand the thermodynamic aspects, compared to the corresponding caloric curve for the isothermal sphere. 

\subsubsection{Singularities in caloric curve of the ss-OAFP model}

\begin{figure}[H]
	\centering
	\begin{tikzpicture}
	\begin{axis}[width=6cm,height=7cm,
	grid=major,xlabel=\Large{$\eta^\text{(kin)}$},ylabel=\Large{$1/\Lambda$},xmin=0,xmax=3,ymin=-0.3,ymax=0,legend pos=south east]
	\addplot [color = red ,mark=no,thick,solid ] table[x index=2, y index=1]{R_InvLamb_INVnTkin_INVnTloc_INVnTcon.txt};
	\node[above,black] at (0.55,-3e-2) {$R_\text{M}=0$};
	\node[above,black] at (1.6,-0.22) {$R_\text{M}\to\infty$};
	\node[above,black] at (2.7,-4e-2) {\Large{\textbf{(a)}}};
	\end{axis}
	\end{tikzpicture}\hspace{0.3cm}
	\begin{tikzpicture}
	\begin{axis}[width=6cm,height=7cm,
	grid=major,xlabel=\Large{$\eta^\text{(con)}$},xmin=0,xmax=3,ymin=-0.3,ymax=0,legend pos=south east]
	\addplot [color = orange ,mark=no,thick,solid ] table[x index=4, y index=1]{R_InvLamb_INVnTkin_INVnTloc_INVnTcon.txt};
	\node[above,black] at (0.55,-3e-2) {$R_\text{M}=0$};
	\node[above,black] at (0.5,-0.225) {$R_\text{M}\to\infty$};
	\node[above,black] at (2.7,-4e-2) {\Large{\textbf{(b)}}};
	\end{axis}	
	\end{tikzpicture}
	
	\vspace{0.5cm}
	
	\begin{tikzpicture}
	\begin{axis}[width=6cm,height=7cm,
	grid=major,xlabel=\large{$\eta^\text{(loc)}$},ylabel=\Large{$1/\Lambda$},xmin=0,xmax=3,ymin=-0.3,ymax=0,legend pos=south east]
	\addplot [color =blue ,mark=no,thick,solid ] table[x index=3, y index=1]{R_InvLamb_INVnTkin_INVnTloc_INVnTcon.txt};
	\node[above,black] at (0.55,-3e-2) {$R_\text{M}=0$};
	\node[above,black] at (2.5,-0.20) {$R_\text{M}\to\infty$};
	\node[above,black] at (2.7,-4e-2) {\Large{\textbf{(c)}}};
	\end{axis}
	\end{tikzpicture}\hspace{0.3cm}
	\begin{tikzpicture}
	\begin{axis}[width=6cm,height=7cm,
	grid=major,xlabel=\large{$\eta^\text{(loc)}$},xmin=2.4,xmax=2.8,ymin=-0.29,ymax=-0.22,legend pos=south east]
	\addplot [color = blue ,mark=no,thick,solid ] table[x index=3, y index=1]{R_InvLamb_INVnTkin_INVnTloc_INVnTcon.txt};
	\node[above,black] at (2.55,-0.23) {$R_\text{M}\to\infty$};
	\node[above,black] at (2.75,-0.23) {\Large{\textbf{(d)}}};
	\end{axis}
	\end{tikzpicture}
	\caption{Caloric curves for the dimensionless- total energy $\Lambda$ and temperatures \textbf{(a)} $\eta^\text{(kin)}$, \textbf{(b)} $\eta^\text{(con)}$ and \textbf{(c)}  $\eta^\text{(loc)}$. \textbf{(d)} Magnification of \textbf{(c)} around the turning point. All the curves start at $(0,0)$ that corresponds to $R_\text{M}=0$. For graphing, the ordinates are a reciprocal of $\Lambda$, hence the tangents to the curves represent the sign of $C_\text{V}$.}
	\label{Fog.R_InvLamb_INVnTkin_INVnTloc_INVnTcon}
\end{figure}
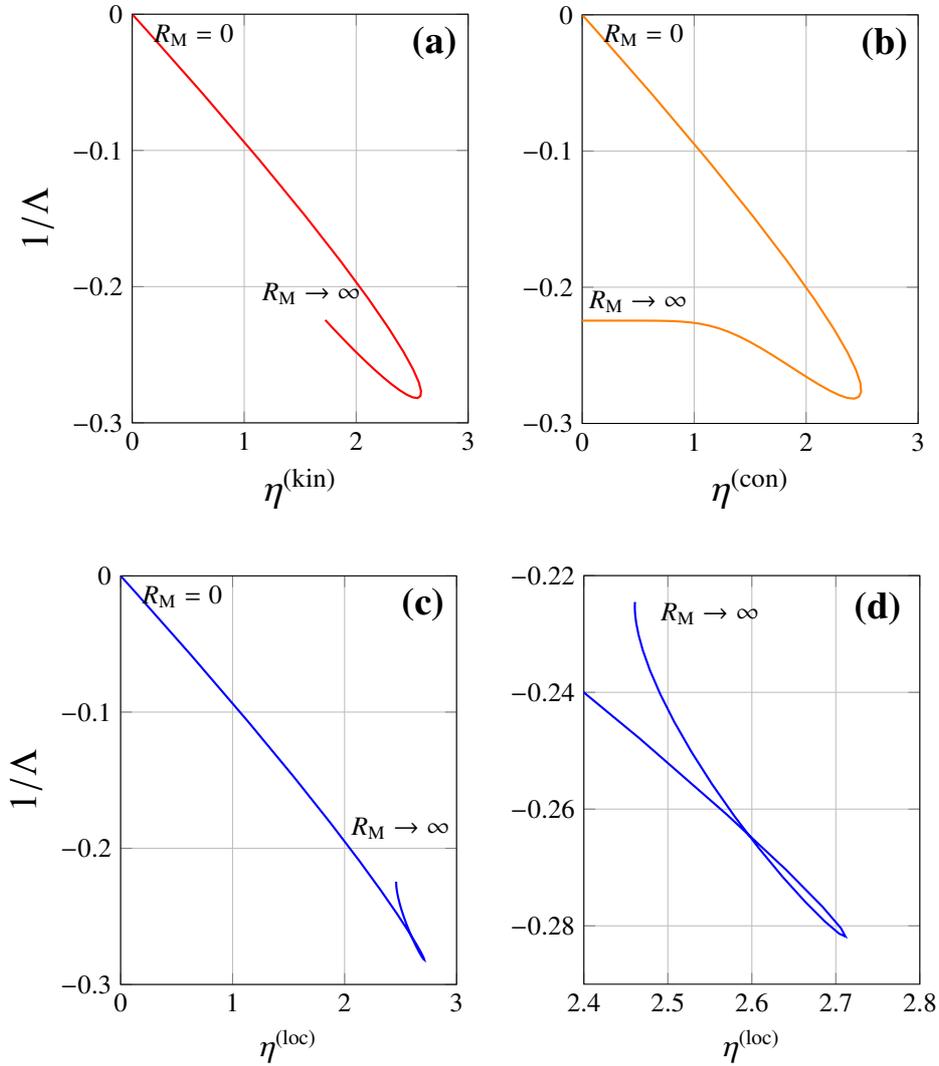

The heat capacity of the ss-OAFP model has singularities at small radii, which differs from the case of the isothermal sphere. Figure \ref{Fog.R_InvLamb_INVnTkin_INVnTloc_INVnTcon} depicts the caloric curves for the normalized energy $\Lambda$ against the normalized inverse temperatures $\eta^\text{(kin)}$, $\eta^\text{(con)}$ and $\eta^\text{(loc)}$. All the caloric curves show negative heat capacity at small radii while they have different characteristics at larger radii beyond the turning points. The turning points of the caloric curves (Table \ref{table:C-curve_Tuning}) occur very close to each other in radius ($R_M\approx18\sim 24$ for $C_\text{V}\to\infty$ and $R_\text{M}\approx25$ for $C_\text{V}=0$ ). Perhaps, this is consistent with the nature of gravothermal instability \emph{if}\footnote{As we discuss in Section \ref{sec:cause_nCv}, the cause of negative specific heat capacity is the relation between heat- and stellar flows under collisionless limit in deep potential well, especially the potential well is essential. Hence, the closeness of the sigularities in radius may occur if one assumes the wall does not affect the structure of potential well.} the negative heat capacity of the core originates from the core being nearly self-gravitating, independent of the halo or being in micro-canonical ensemble \citep{Lynden_Bell_1968,Thirring_1970,Lynden_Bell_1977} \footnote{For an isolated self-gravitating system, the Virial theorem states $2KE+PE=0$, so $E_\text{tot}=-$KE \citep[e.g.][]{Heggie_1988,Binney_2011}. Hence, the kinetic heat capacity is $C_{V}\propto E_\text{tot}/\text{KE}<0$. As \cite{Thirring_1970} originally pointed out, the energy range of a system that has negative heat capacity in micro-canonical ensemble corresponds to the same energy band of the system that undergoes phase transition in canonical ensemble. Simply, canonical ensemble applies to the isothermal sphere at radii smaller than $R_\text{M}=8.9$ and micro-canonical ensemble at radii smaller than $R_\text{M}=34.9$ (See Table \ref{table:C-curve_Tuning} for the value of radii.). The heat capacity is positive on $0<R_\text{M}<8.9$ and negative on $8.9<R_\text{M}<34.9$.}. This implies that regardless of kinds of walls (adiabatic or thermal) and ensemble (micro-canonical or canonical), the wall less affects any dynamics of the core. Hence, the singular behaviors in the heat capacity could occur very close to each other in radius \emph{beyond} the radii $R_\text{M}=4.31$ and $R_\text{M}=3.31$ at which the heat- and stellar fluxes reach their maximum. This property is the distinct difference from the isothermal sphere. In the case of the isothermal sphere, the radii at which the singularities occur are relatively different from each other at small radii ($R_\text{M}=34.9$ for $C_\text{V}\to\infty$ and $R_\text{M}=8.9$ for $C_\text{V}=0$ ).

\begin{table*}\centering
	\ra{1.3}
	\begin{tabular}{@{}l|cccc|cccc@{}}\toprule
		&$C_\text{V}\to\infty$ &    &&         & $C_\text{V}=0$ &  & &\\ 
		\midrule
		& $\eta$ & $\Lambda$&$R_\text{M}$ & $D(0)/D(R_\text{M})$ &$\eta$& $\Lambda$ &$R_\text{M}$ & $D(0)/D(R_\text{M})$  \\ 
		\midrule
		Kinetic & 2.580  &-3.599  &  19.2   &  36.6  &2.540& -3.548&25.4&74.85\\
		Local   & 2.712  &-3.549  &  24.1   &  65.8  &2.711& -3.548&25.4&74.85\\
		Thermodynamic & 2.492  &-3.633  &  17.9   &  30.3  &2.418&  -3.548&25.4&74.85\\
		\midrule
		IS      &        &-0.335  &  34.4   &  709   &2.52&          &8.99   &32.1\\
		\bottomrule
	\end{tabular}\\
	\caption{First turning points of caloric curves at small radii. The data for IS are the corresponding values for the isothermal sphere adapted from \citep{Antonov_1962, Lynden_Bell_1968, Padmanabhan_1989, Chavanis_2002_iso}.}
	\label{table:C-curve_Tuning}
\end{table*}

\section{Cause of negative heat capacity}\label{sec:cause_nCv}

Since the ss-OAFP model is at a non-equilibrium state, the self-gravity of the model does not cause negative heat capacity unlike (nearly) isolated self-gravitating systems at a state of equilibrium. The present section details possible cause of the negativeness especially focusing on the core of the model, accordingly only the kinetic heat capacity is discussed for simplicity. First, Section \ref{sec:cause_nCv_center} introduces a simple analytical method to discuss the heat capacity based on the Virial and total energy of the model. This method shows that the shallow potential well at the center of the isothermal sphere causes positive kinetic heat capacity. On one hand, it also shows that the negative heat capacity originates from the deep potential well (or large scaled escape energy)  at the center of the core in the ss-OAFP model. Lastly, Section \ref{cause_nCV_comp} shows that the high-temperature- and collisionless limits of the core drive the negative heat capacity on relaxation time scale by comparing the present work to existing works. 

\subsection{Virial and total energy to discuss heat capacity at the center of the core}\label{sec:cause_nCv_center}

The heat capacity at the center of the ss-OAFP model can be discussed by the Virial and total energy $E_\text{tot}$ (equation \eqref{Eq.Etot_ss})
\begin{subequations}
	\begin{align}
	&\mathcal{V}\equiv-2\text{KE}-\text{PE}=-\int^{R_\text{M}}_{0}\,\text{d}R\int^{0}_{\Phi}\,\text{d}E\,\Omega_\text{mic}(E,\Phi)\,R^{2}(E)\,(E-\Phi)^{3/2},\label{Eq.Virial}\\
	&E_\text{tot}=\int^{R_\text{M}}_{0}\,\text{d}R\int^{0}_{\Phi}\,\text{d}E\,\Omega_\text{can}(E,\Phi)\,R^{2}(E)\,(E-\Phi)^{3/2},\label{Eq.E_tot}
	\end{align}
\end{subequations}
where 
\begin{subequations}
	\begin{align}
	&\Omega_\text{mic}(E,\Phi)=2F(E)+\frac{\Phi}{3}\frac{\,\text{d}F}{\,\text{d}E},\label{Eq.intgerand kernal_mic}\\
	&\Omega_\text{can}(E,\Phi)=F(E)+\frac{\Phi}{3}\frac{\,\text{d}F}{\,\text{d}E}.\label{Eq.intgerand kernal_can}
	\end{align}
\end{subequations}
Equations \eqref{Eq.intgerand kernal_mic} and \eqref{Eq.intgerand kernal_can} present a simple analytical method to determine the sign of heat capacity in the core. Since the present focus is the center of the cores of star clusters, one may assume stellar DF obeys Maxwellian DF with escape energy $\chi_\text{esc}$; the dimensionless DF reads $F(E)=\exp[\chi_\text{esc}E]$. Then, $\Omega_\text{mic}$ and $\Omega_\text{con}$ reduce to
\begin{subequations}
	\begin{align}
	&\Omega_\text{mic}(E,\Phi)=e^{\chi_\text{esc}E}\left(2+\frac{\chi_\text{esc}}{3}\Phi\right),\label{Eq.Omega_mic}\\
	&\Omega_\text{con}(E,\Phi)=e^{\chi_\text{esc}E}\left(1+\frac{\chi_\text{esc}}{3}\Phi\right).\label{Eq.Omega_con}
	\end{align}
\end{subequations}
Change in the sign of heat capacity depends on
\begin{subequations}
	\begin{align}
	\Phi^\text{(mic)}=-\frac{6}{\chi_\text{esc}},\label{Eq_crit_neg_PEKE_mic}\\
	\Phi^\text{(can)}=-\frac{3}{\chi_\text{esc}}.\label{Eq_crit_neg_PEKE_can}
	\end{align}
\end{subequations}
These equations may be also related to the singularities in heat capacity.\footnote{Equation \eqref{Eq_crit_neg_PEKE_mic} corresponds to the Virial being zero, that is, the thermal pressure at the adiabatic wall is zero. This determines the upper limit radius beyond which micro-canonical ensemble can not apply to a self-gravitating system if the system is at a state of equilibrium.  Equation \eqref{Eq_crit_neg_PEKE_can} corresponds to $E_\text{tot}$ being zero, that is, $E_\text{tot}$ is dominated by PE. This is a minimum condition that an equilibrium self-gravitating system does not exist in canonical ensemble. Under proper thermodynamic limit, the dimensionless total energy $\Lambda$ reaches an extremum (singular point) as $R_\text{M}$ increases since, for large $R_\text{M}$, $E_\text{tot}$ is proportional to $M^2 G/R_\text{M}$. At radii larger or smaller than the singular point, the sign of $C_\text{V}$ possibly changes.} 

\begin{figure}
	\centering
	\begin{tikzpicture}
	\begin{semilogxaxis}[width=11cm,height=5cm,
	grid=major,ylabel=\Large{$-\frac{\text{3PE}}{\chi_\text{esc}\text{KE}}$},xmin=1e-1,xmax=1e18,ymin=0,ymax=1.2,legend pos=south east]
	\addplot [color = red ,mark=no,thick,solid ] table[x index=0, y index=1]{R_PeKeRatio_regVirial.txt};
	\end{semilogxaxis}
	\end{tikzpicture}
	\caption{Ratio of -PE/KE to $\chi_\text{esc}/3$.}
	\label{Fig.R_PeKeRatio}
\end{figure}
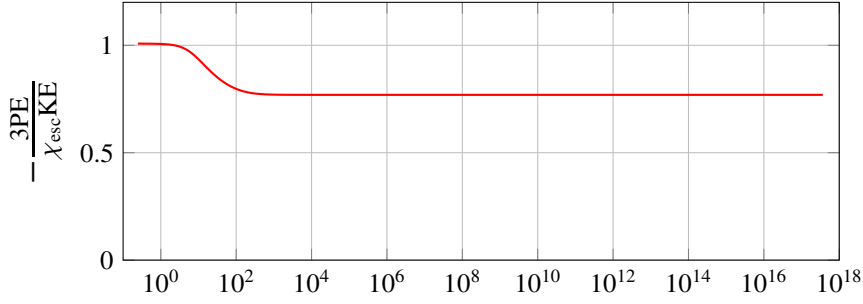

The present section relies on the approximation $F=e^{\chi_\text{esc}E}$, hence one needs to test the approximation. One still can use equations \eqref{Eq.intgerand kernal_mic} and \eqref{Eq.intgerand kernal_can} with $\chi_\text{esc}=13.88$ for the ss-OAFP model. This is since the reference DF $F_\text{o}(E)$ well fits the exponential of $\chi_\text{esc}E$ on $-1<E\lessapprox-0.5$ and $\exp[-\chi_\text{esc}0.5]$ contributes to the integrals (equations \eqref{Eq.Virial} and \eqref{Eq.E_tot}) only by a small fraction ($\sim 1\times10^{-3}$). Also, Figure \ref{Fig.R_PeKeRatio} shows the ratio $-\text{PE}/\text{KE}$ regularized by $\chi_\text{esc}/3$ and it is a direct graphical representation for Equation \ref{Eq.intgerand kernal_mic}. The value of $-3\text{PE}/\text{KE}\chi_\text{esc}$ reaches unity at $R_\text{M}\approx 0$, which validates the approximation of the DF to $e^{\chi_\text{esc}E}$ at the center of the core.
The rest applies equations \eqref{Eq.Omega_mic} and \eqref{Eq.Omega_con} to the kinetic heat capacities of isothermal sphere (Section \ref{sec:pCv_center_ISO}) and the ss-OAFP model (Section \ref{sec:nCv_center_OAFP}).

\subsubsection{ Positive kinetic heat capacity at the center of the isothermal sphere}\label{sec:pCv_center_ISO}

First the isothermal sphere ($\chi_\text{esc}=1$) is discussed whose kinetic heat capacity is positive at the center of the core due to the shallow potential well. The relation between the potential well and $R_\text{M}$ is known \citep[see e.g.][]{Lynden_Bell_1968,Padmanabhan_1989}. As $R_\text{M}\to\infty$ the potential well deepens like $\beta\phi(0)=-2\ln[R]-2$ due to the relationship $\beta\phi(0)=-\Phi(R_\text{M})-M/R_\text{M}$. Hence, as $R_\text{M}\to\infty$ the potential well gets deeper and the total energy decreases. On one hand, for small $R_\text{M}$ such as $R_\text{M}<<1$, $\beta\phi(0)=-R_\text{M}^{2}/6$. Hence, even if $R_\text{M}=1$, $E_\text{tot}$ and $\mathcal{V}$ are necessary positive. The energy $E_\text{tot}$ increases monotonically with $R_\text{M}$\footnote{Controlling the value of $E_\text{tot}$ and KE by $R_\text{M}$ is essentially based on the dependence of the normalized energy and temperature on $R_\text{M}$. As shown in Section \ref{sec:Reg_C-curve} the normalized quantities implicitly depend on radius $R_\text{M}$ on caloric curve even if the dependence does not appear explicitly in Legendre transformation. In other words, we expect that controlling $R_\text{M}$ corresponds to changing the state of system from a non-equilibrium state to another non-equilibrium state, which is an analogy of tracing infinitely possible equilibria for equilibrium self-gravitating system.} (due to the integrand being positive). Also, the kinetic energy  monotonically increases with $R_\text{M}$ at all radii (since local kinetic energy is always positive). The energies KE and $E_\text{tot}$ are zero at $R_\text{M}=0$ and are made in dimensionless form by the same factor (Section \ref{sec:Reg_C-curve})\footnote{The factor $M^{2}/R_{M}$ behaves like $R_{M}^{9}/900$ as $R_\text{M}\to0$ which changes the increasing behavior of $KE$ and $E_\text{tot}$ into decreasing functions with small increasing $R_\text{M}$ though, $M^{2}/R_{M}$ still changes monotonically and much rapidly compared to $E_\text{tot}$ and KE at $R_\text{M}\sim0$ hence it does not alter the sign of heat capacity.}. As a result, the kinetic heat capacity is positive in the core for the isothermal sphere due to the shallow potential well. This is the case that we consider the core of the isothermal sphere as an `ordinary' ideal gas (since the m.f. potential less affects the state of core).

\subsubsection{Negative heat capacity at the center of the ss-OAFP model}\label{sec:nCv_center_OAFP}

In the case of the ss-OAFP model, the negative kinetic heat capacity at the center of the core is caused by the deep potential well. Thanks to the self-similarity of the ss-OAFP model, the potential $\Phi$ at $R_\text{M}\approx0$ is not related to the wall radius $R_\text{M}$ unlike the isothermal sphere. Hence, one can discuss the heat capacity based only on the local property of the ss-OAFP model. The known approximation, $\Phi(R_\text{M}\approx 0)=-1+R_\text{M}^{2}/6$, infers that $E_\text{tot}$ and $\mathcal{V}$ are negative at $R_\text{M}<2.17$ and $R_\text{M}<1.85$. Accordingly, $E_\text{tot}$ decreases with $R_\text{M}$ at small radii (due to the integrand being negative).  Also, the kinetic energy increases at all radii as $R_\text{M}$ increases. In the same way as the isothermal sphere, the energies KE and $E_\text{tot}$ are zero at $R_\text{M}=0$ and are made in dimensionless form by the same factor (Section \ref{sec:Reg_C-curve}). Hence, the ss-OAFP model must have a negative heat capacity at small radii ($R_\text{M}\lessapprox2.2$) in the core. One may recall  $\Phi(R_\text{M}=0)$ is set to $-1$ for numerical integration of the ss-OAFP system. Yet, through equations \eqref{Eq.intgerand kernal_mic} and \eqref{Eq.intgerand kernal_can}, a large $\chi_\text{esc}$-value is equivalent to a deep potential well. Hence, the negative heat capacity is the consequence of the deep potential well. This is the case that the present work considers the core of the ss-OAFP model behaves as an `exotic' ideal gas.

\subsection{The cause of negative heat capacity compared to the previous works}\label{cause_nCV_comp}

The present section compares the negative heat capacity of the core of the ss-OAFP model to that of a simple model (Section \ref{sec:simple_nCv}) and  more realistic model (Section \ref{sec:real_nCv}) to discuss the cause of the negativity.

\subsubsection{A simple understanding of the negative heat capacity in the core of the ss-OAFP model}\label{sec:simple_nCv}

For the ss-OAFP model, a small number of stars is trapped in the deep potential well due to the high escape energy (Figure \ref{Fig.R_VescTOT_VescCON}) while the spatial profile of stars in the halo is time-independent.\footnote{As $R_\text{M}>>1$, the halo density $\rho(r,t)$ is independent of the dynamics in the core or relaxation evolution in self-similar analysis, which provides $\text{d}\rho(r,t)/\text{d}t=0$, hence $D(R)\propto R^{-\alpha}$ and $\rho_\text{t}\propto r_{t}^{-\alpha}$.} Hence, the trapped stars in the core may keep staying in the core potential well. The stars more tightly bound each other every time the core loses stars (particles) and kinetic- energy through the heat- and -stellar fluxes due to the conservation of energy\footnote{One may recall the failure of the Bohr model in which electron releases electromagnetic radiation due to the acceleration and deeply penetrates into the potential well around an ion.}. This means if one encloses the stars in the core by an adiabatic wall of small $R_\text{M}$ they may behave like particles interacting via short-range pair potential  \citep{Posch_1990, Posch_2005, Thirring_2003} though, this is not the case.  For such systems, the majority of energy bands (levels) is available to particles meaning the particles can be well mixed and the negative heat capacity occurs to only limited energy band that corresponds to a phase transition. On one hand, the present work deals with only the collapsed-state. Also, the energy band that shows negative heat capacity is broad in the sense that it covers at least the energy band that is available to particles in the potential well. 

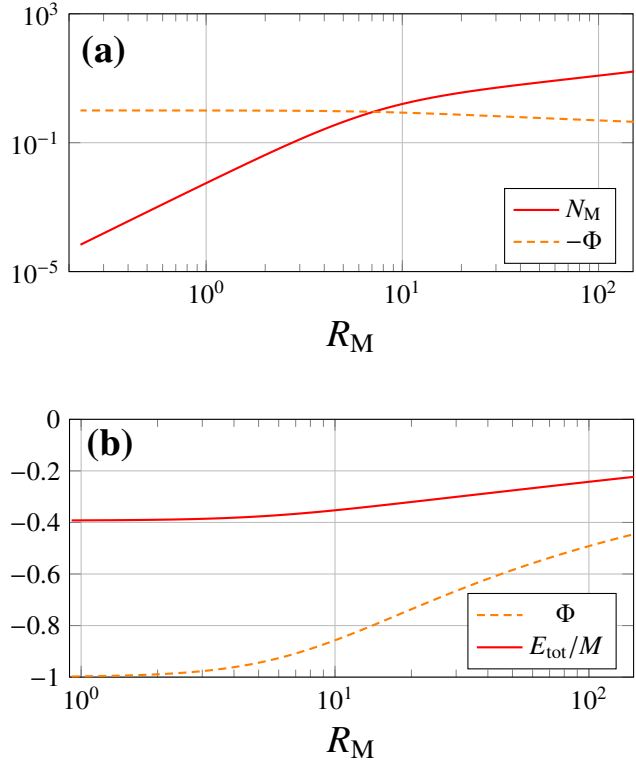
\begin{figure}[H]
	\centering
	\begin{tikzpicture}
	\begin{loglogaxis}[width=9cm,height=5cm,
	grid=major,xlabel=\Large{$R_\text{M}$},xmin=2e-1,xmax=150,ymin=1e-5,ymax=1e3,legend pos=south east]
	\addplot [color = red ,mark=no,thick,solid ] table[x index=0, y index=1]{R_M_nPhi.txt};
	\addlegendentry{\normalsize{$N_\text{M}$}};
	\addplot [color = orange,mark=no,thick,densely dashed ] table[x index=0, y index=2]{R_M_nPhi.txt};
	\addlegendentry{\normalsize{$-\Phi$}};
	\node[above,black] at (0.3,1e1) {\Large{\textbf{(a)}}};
	\end{loglogaxis}
	\end{tikzpicture}
	
	\vspace{0.5cm}
	
	\begin{tikzpicture}
	\begin{semilogxaxis}[width=9cm,height=5cm,
	grid=major,xlabel=\Large{$R_\text{M}$},xmin=9e-1,xmax=150,ymin=-1,ymax=-0.0,legend pos=south east]
	\addplot [color = orange ,mark=no,thick,densely dashed ] table[x index=0, y index=1]{R_Phi_normEtot.txt};
	\addlegendentry{\normalsize{$\Phi$}};
	\addplot [color = red,mark=no,thick,solid ] table[x index=0, y index=2]{R_Phi_normEtot.txt};
	\addlegendentry{\normalsize{$E_\text{tot}/M$}};
	\node[above,black] at (1.3,-0.2) {\Large{\textbf{(b)}}};
	\end{semilogxaxis}
	\end{tikzpicture}
	\caption{Dimensionless m.f. potential $\Phi$ compared to \textbf{(a)} total number of stars for the ss-OAFP model enclosed by an adiabatic wall at $R_\text{M}$ and compared to \textbf{(b)} normalized mean energy per unit mass of the ss-OAFP model.}
	\label{Fig.R_Phi_normEtot}
\end{figure}

The situation for stars in the core of the ss-OAFP model is alike that for collisionless particles in a potential well that reported negative heat capacity \citep{Thirring_2003,Carignano_2010}. For the core being enclosed by an adiabatic wall, the initial conditions of dynamical evolution are not important since the core is a well-relaxed (non-)equilibrium state. The two key points here are that the temperature at the center is high (Figure \ref{Fig.R_nTkin_nTloc_nTcon}) and that the probability to find stars in the core is low. For the latter, the total number $N_\text{M}(\equiv M/m)$ of stars in the core is a small fraction e.g. $N_\text{M}\approx 5.6\times10^{-3}$ at $R_\text{M}=1$ and $N_\text{M}\approx 1$ at the radius at which the flattening in $\Phi$ ceases (Figure \ref{Fig.R_Phi_normEtot} \textbf{(a)}). Hence, a proper zeroth-order approximation is collisionless limit. In this limit, the core behaves like a collisionless ideal gas as typically assumed for the isothermal sphere \citep[e.g.][]{Katz_1978}. This means the stars behave as if they were non-interacting particles traveling only under the effect of m.f. potential $\Phi$. Figure \ref{Fig.R_Phi_normEtot} \textbf{(b)} shows the potential well $\Phi(R_\text{M})$ and mean total energy per unit mass $E_\text{tot}/M$.  The stars in the core can stay in the deep potential well to develop the core-collapse, however due to the high temperature (kinetic energy) some of stars need to spill out of the potential well. The corresponding thought experiment is to 'expand' the adiabatic wall\footnote{We still follow the basic idea of the infinitely possible equilibria used for caloric curves (Section \ref{sec:Reg_C-curve}) and the 'expand' means that one does move on to another non-equilibrium state at larger radius rather than physically expand the wall itself.}. Then from Figure \ref{Fig.R_Phi_normEtot}\textbf{(b)}, the total kinetic energy per unit mass obviously decreases while the total energy increases with radius. Hence, the negative heat capacity is the outcome of the deep potential well together with the high temperature and low total number of stars in the core. The former is related only to the structure of system hence the time scale does not matter while the latter is peculiar to non-equilibrium systems on relaxation time scales to hold the negative heat capacity. This graphical method \citep{Thirring_2003, Carignano_2010} would be the simplest way to understand the negative heat capacity in the present case. The present result adds to their discussions the importance of heat- and particle- flows to 'keep' negative heat capacity on long (relaxation) time scale, in short, the significance of non-equilibrium state.  

\subsubsection{Comparison to an existing realistic model}\label{sec:real_nCv}

Lastly, we compare the physical state of the core of the ss-OAFP model to the results of $N$-body simulations \citep{Komatsu_2010,Komatsu_2012} executed under physical conditions similar to the core. Strictly speaking, systems similar to the ss-OAFP model do not exit not only in nature but also as a result of numerical $N$-body simulation\footnote{The reason why one is hard to achieve the state of the ss-OAFP model using $N$-body simulation is that it is not easy to achieve the 'complete' core collapse other than using continuum models. Not only the effect of binary stars stops core collapse but also large $N\approx10^5$ costs unfeasible CPU to achieve the complete collapse. Hence, one needs to find a similarity of the ss-OAFP model to self-gravitating system of less $N$ stars. } since complete core-collapse itself is a mathematical concept. Yet, some features of complete core-collapse should appear at the early stage of core collapse as shown in \citep{Cohn_1980}. The conditions similar to the core of the ss-OAFP model are that a self-gravitating system of small $N$ particles must be enclosed by a wall undergoing a core-collapse but losing kinetic-energy and stars toward the outside of the wall due to heat- and stellar- fluxes. Also, the system must be large enough to form core-halo structure to make the fluxes occur. \citep{Komatsu_2010, Komatsu_2012} embodies such conditions by using a $N$-body simulation for $N=125\sim250$. To achieve the condition, they used a partially permeable wall through which the evaporation (escape) rate of particles from the system can change by controlling the escape energy, that is, the degree of permeability.

Our result can reach the energy domain that the results of \citep{Komatsu_2010, Komatsu_2012} could not achieve while the both results show many common physical features. \cite{Komatsu_2012} obtained a Maxwellian-like velocity distribution function, and showed not only negative heat capacity but also greater velocity gradient with increasing time \citep[like][]{Cohn_1980}, together with density and dispersion well correlating (like Figure \ref{Fig.R_Phi_Vdis_HeatF_DensF_Tr}\textbf{(a)}). On one hand, \cite{Komatsu_2010} employed stellar polytropes to characterize the non-equilibrium state of their model using the same method. They found the polytrope of $m\sim9$ is initially close to their model with a negative value of $\Lambda$ but it begins to deviate from the model with decreasing $\Lambda$ as the collapse proceeds. They insisted the deviation occurred because the kinetic temperature is not suitable to describe the non-equilibrium state. Yet, it would less matter. Smaller $\Lambda$ means the the core size is relatively large and close to the wall radius. This corresponds to, in the present model, radius $R_\text{M}$ approaches the center of the core where the kinetic temperature can be reasonably defined due to the frequent relaxation events. The reason the deviation occurred would be because the polytrope itself is not proper to describe the system with negative heat capacity. The caloric curve for the polytropes of $m>5$ spirals as $\Lambda$ decreases, showing marginal instabilities (or successive instability with increasing radius) \citep{Chavanis_2002_poly,Chavanis_2003}. This situation corresponds with collapsed state with large $R_\text{M}$ which does not exist due to the instability.  On one hand, the caloric curve for the ss-OAFP model provides only negative heat capacity at small $\Lambda$ (or monotonically increasing $1/\eta^{(kin)}$ with decreasing $\Lambda$). This well matches the qualitative nature of the curve reported in \citep[Figure 5 in][]{Komatsu_2010}. 

\section{Conclusion} \label{sec:Conclusion}

In the present paper we aimed at showing the basic physical features of the ss-OAFP model focusing on the core of the model and negative heat capacity, based on our previous numerical results for the pre-collpase solution to the ss-OAFP equation \citep{Ito_2020_1}. We first discussed the local and global properties of the ss-OAFP model. The model shows similar properties compared to the self-similar conductive gaseous model though, we found the equation of state in the core is local ideal gas $p=1.0\rho/\chi_\text{esc}$ while it is polytropic $p=0.5\rho^{1.1}/\chi_\text{esc}$ at large radii. Since the the center of the core can be described by the state of the polytropic index $m=177$, it shows the incompleteness of Maxwellain DF even at the center. We also showed the negative heat capacity at constant volume of the model by constructing regularized quantities. A unique feature of the ss-OAFP model originates from the model being at a nonequilibrium state. Although the core is still well-relaxed state, it can achieve negative heat capacity since stars in the core can behave like collisionless particles due to high temperature and low total number of stars in its deep potential. This cause of negative heat capacity is different from equilibrium self-gravitating systems such as the isothermal sphere for which negative heat capacity occurs as a result of phase transition or when the system is large enough to be isolated from the ambient stars/gas in the sense that the Virial reaches zero. The Virial of the ss-OAFP model is not zero, rather positive. Also, the ss-OAFP model is not related to the cases in which equilibrium systems can show negative heat capacity because of the initial conditions. The present analysis reemphasizes that the negative heat capacity can be peculiar to self-gravitating non-equilibrium systems in addition to inhomogeneous equilibrium systems. 

The present work revealed the similarity and difference between the cores of the isothermal sphere and ss-OAFP model. Considering their similarity in structure of the cores, we may also expect that there is a relation between the King model and ss-OAFP model after the latter is properly truncated in energy space. Hence, in the follow-up paper \citep{Ito_2020_3} we will discuss an application of an energy-truncated ss-OAFP model to the structural profiles of Galactic globular clusters compared to the results of the King model.

\begin{appendices}
\renewcommand{\theequation}{\Alph{section}.\arabic{equation}}

\section{Regularization of total energy}\label{App.norm_Etot}

The present appendix explains the relationship between the regularization of $E_\text{tot}$ and Poisson equation for the ss-OAFP model. A proper integral of Poisson equation in inverse form \citep{Ito_2020_1} with respect to $R$ reads
\begin{align}
R^{2}\frac{\text{d}\Phi}{\text{d}R}=M,
\end{align}
Since the ss-OAFP model has power law boundary condition $\Phi\propto R^{-\alpha}$ as $R\to\infty$, one can employ the identity $\frac{\text{d}\Phi}{\text{d}R}=-\alpha\Phi/R$. Accordingly, the Poisson equation as large $R$ reduces to
\begin{align}
\Phi=-\frac{M}{R\alpha}.
\end{align}
The total energy $E_\text{tot}$ at $R\to \infty$ is obviously proportional to $M$ and the domain of $E_\text{tot}$ is $(\Phi,0)$. Hence, a proper change of variable makes the factor $(E-\Phi)$ be proportional to $\Phi$. As a result, 
\begin{align}
E_\text{tot}\propto\frac{M^{2}}{R}.
\end{align} 
Although one may choose different quantities to regularize $E_\text{tot}$, we focus on heat capacity at constant volume. This means total number $N$ and volume $V$ are constant,  accordingly total mass $M$ and radius $R$ are constant (under Legendre transformation). Hence, the regularization of $E_\text{tot}$ in the present work originates from the characteristics of the Poisson equation.

\end{appendices}
\section*{Acknowledgements}
The present work is partial fulfillment of the degree of Philosophy at CUNY graduate center.
\vspace{1cm}

\bibliographystyle{elsarticle-harv} 
\bibliography{science}

\begin{thebibliography}{54}
\expandafter\ifx\csname natexlab\endcsname\relax\def\natexlab#1{#1}\fi
\expandafter\ifx\csname url\endcsname\relax
  \def\url#1{\texttt{#1}}\fi
\expandafter\ifx\csname urlprefix\endcsname\relax\def\urlprefix{URL }\fi

\bibitem[{Aguado and Jarrold(2011)}]{Aguado_2011}
Aguado, A., Jarrold, M.~F., may 2011. Melting and freezing of metal clusters.
  Annual Review of Physical Chemistry 62~(1), 151--172.
\newline\urlprefix\url{https://doi.org/10.1146%2Fannurev-physchem-032210-103454}

\bibitem[{Antonov(1962)}]{Antonov_1962}
Antonov, V., 1962. Most probable phase distribution in spherical star systems
  and conditions for its existence. Vest. Leningrad Univ. 7, 135.

\bibitem[{Binney and Tremaine(2011)}]{Binney_2011}
Binney, J., Tremaine, S., 2011. Galactic Dynamics. Princeton university press.

\bibitem[{Borderie and Frankland(2019)}]{Borderie_2019}
Borderie, B., Frankland, J., mar 2019. Liquid{\textendash}gas phase transition
  in nuclei. Progress in Particle and Nuclear Physics 105, 82--138.
\newline\urlprefix\url{https://doi.org/10.1016%2Fj.ppnp.2018.12.002}

\bibitem[{Brilliantov et~al.(2018)Brilliantov, Formella, and
  Pöschel}]{Brilliantov_2018}
Brilliantov, N.~V., Formella, A., Pöschel, T., feb 2018. Increasing
  temperature of cooling granular gases. Nature Communications 9~(1).
\newline\urlprefix\url{https://doi.org/10.1038%2Fs41467-017-02803-7}

\bibitem[{Carignano and Gladich(2010)}]{Carignano_2010}
Carignano, M.~A., Gladich, I., jun 2010. Negative heat capacity of small
  systems in the microcanonical ensemble. {EPL} (Europhysics Letters) 90~(6),
  63001.
\newline\urlprefix\url{https://doi.org/10.1209%2F0295-5075%2F90%2F63001}

\bibitem[{Chavanis(2002{\natexlab{a}})}]{Chavanis_2002_iso}
Chavanis, P.~H., jan 2002{\natexlab{a}}. Gravitational instability of finite
  isothermal spheres. Astronomy {\&} Astrophysics 381~(1), 340--356.
\newline\urlprefix\url{https://doi.org/10.1051%2F0004-6361%3A20011438}

\bibitem[{Chavanis(2002{\natexlab{b}})}]{Chavanis_2002_poly}
Chavanis, P.~H., may 2002{\natexlab{b}}. Gravitational instability of
  polytropic spheres and generalized thermodynamics. Astronomy {\&}
  Astrophysics 386~(2), 732--742.
\newline\urlprefix\url{https://doi.org/10.1051%2F0004-6361%3A20020306}

\bibitem[{Chavanis(2003)}]{Chavanis_2003}
Chavanis, P.~H., mar 2003. Gravitational instability of isothermal and
  polytropic spheres. Astronomy {\&} Astrophysics 401~(1), 15--42.
\newline\urlprefix\url{https://doi.org/10.1051%2F0004-6361%3A20021779}

\bibitem[{Claydon et~al.(2019)Claydon, Gieles, Varri, Heggie, and
  Zocchi}]{Claydon_2019}
Claydon, I., Gieles, M., Varri, A.~L., Heggie, D.~C., Zocchi, A., may 2019.
  Spherical models of star clusters with potential escapers. Monthly Notices of
  the Royal Astronomical Society 487~(1), 147--160.
\newline\urlprefix\url{https://doi.org/10.1093%2Fmnras%2Fstz1109}

\bibitem[{Cohn(1980)}]{Cohn_1980}
Cohn, H., dec 1980. Late core collapse in star clusters and the gravothermal
  instability. The Astrophysical Journal 242, 765.
\newline\urlprefix\url{https://doi.org/10.1086%2F158511}

\bibitem[{D.~Lynden-Bell and Royal(1968)}]{Lynden_Bell_1968}
D.~Lynden-Bell, R.~W., Royal, A., feb 1968. The gravo-thermal catastrophe in
  isothermal spheres and the onset of red-giant structure for stellar systems.
  Monthly Notices of the Royal Astronomical Society 138~(4), 495--525.
\newline\urlprefix\url{http://dx.doi.org/10.1093/mnras/138.4.495}

\bibitem[{de~Vega and S{\'{a}}nchez(2002)}]{de_Vega_2002}
de~Vega, H., S{\'{a}}nchez, N., mar 2002. Statistical mechanics of the
  self-gravitating gas: I. thermodynamic limit and phase diagrams. Nuclear
  Physics B 625~(3), 409--459.
\newline\urlprefix\url{https://doi.org/10.1016%2Fs0550-3213%2802%2900025-1}

\bibitem[{D{\'{e}}Agostino et~al.(1999)D{\'{e}}Agostino, Botvina, Bruno,
  Bonasera, Bondorf, Bougault, D{\'{e}}sesquelles, Geraci, Gulminelli, Iori,
  Neindre, Margagliotti, Mishustin, Moroni, Pagano, and
  Vannini}]{D_Agostino_1999}
D{\'{e}}Agostino, M., Botvina, A., Bruno, M., Bonasera, A., Bondorf, J.,
  Bougault, R., D{\'{e}}sesquelles, P., Geraci, E., Gulminelli, F., Iori, I.,
  Neindre, N.~L., Margagliotti, G., Mishustin, I., Moroni, A., Pagano, A.,
  Vannini, G., apr 1999. Thermodynamical features of multifragmentation in
  peripheral au $\mathplus$ au collisions at 35 a {MeV}. Nuclear Physics A
  650~(3), 329--357.
\newline\urlprefix\url{https://doi.org/10.1016%2Fs0375-9474%2899%2900097-4}

\bibitem[{D{\'{e}}Agostino et~al.(2000)D{\'{e}}Agostino, Gulminelli, Chomaz,
  Bruno, Cannata, Bougault, Gramegna, Iori, Neindre, Margagliotti, Moroni, and
  Vannini}]{D_Agostino_2000}
D{\'{e}}Agostino, M., Gulminelli, F., Chomaz, P., Bruno, M., Cannata, F.,
  Bougault, R., Gramegna, F., Iori, I., Neindre, N.~L., Margagliotti, G.,
  Moroni, A., Vannini, G., feb 2000. Negative heat capacity in the critical
  region of nuclear fragmentation: an experimental evidence of the liquid-gas
  phase transition. Physics Letters B 473~(3-4), 219--225.
\newline\urlprefix\url{https://doi.org/10.1016%2Fs0370-2693%2899%2901486-0}

\bibitem[{Destri and de~Vega(2007)}]{Destri_2007}
Destri, C., de~Vega, H., feb 2007. Dilute and collapsed phases of the
  self-gravitating gas. Nuclear Physics B 763~(3), 309--329.
\newline\urlprefix\url{https://doi.org/10.1016%2Fj.nuclphysb.2006.10.028}

\bibitem[{Einarsson(2004)}]{Einarsson_2004}
Einarsson, B., nov 2004. Conditions for negative specific heat in systems of
  attracting classical particles. Physics Letters A 332~(5-6), 335--344.
\newline\urlprefix\url{https://doi.org/10.1016%2Fj.physleta.2004.09.066}

\bibitem[{Gobet et~al.(2002)Gobet, Farizon, Farizon, Gaillard, Buchet,
  Carr{\'{e}}, Scheier, and Märk}]{Gobet_2002}
Gobet, F., Farizon, B., Farizon, M., Gaillard, M.~J., Buchet, J.~P.,
  Carr{\'{e}}, M., Scheier, P., Märk, T.~D., oct 2002. Direct experimental
  evidence for a negative heat capacity in the liquid-to-gas phase transition
  in hydrogen cluster ions: Backbending of the caloric curve. Physical Review
  Letters 89~(18).
\newline\urlprefix\url{https://doi.org/10.1103%2Fphysrevlett.89.183403}

\bibitem[{Heggie and Stevenson(1988)}]{Heggie_1988}
Heggie, D.~C., Stevenson, D., jan 1988. Two homological models for the
  evolution of star clusters. Monthly Notices of the Royal Astronomical Society
  230~(2), 223--241.
\newline\urlprefix\url{http://dx.doi.org/10.1093/mnras/230.2.223}

\bibitem[{Ingel(2000)}]{Ingel_2000}
Ingel, L.~K., nov 2000. {\textquotedblleft}negative heat
  capacity{\textquotedblright} of stratified fluids. Journal of Experimental
  and Theoretical Physics Letters 72~(10), 527--529.
\newline\urlprefix\url{https://doi.org/10.1134%2F1.1343157}

\bibitem[{{Ito}(2020{\natexlab{a}})}]{Ito_2020_1}
{Ito}, Y., Mar. 2020{\natexlab{a}}. {Self-similar orbit-averaged Fokker-Planck
  equation for isotropic spherical dense clusters (i) accurate pre-collapse
  solution}. arXiv e-prints, arXiv:2003.12196.

\bibitem[{{Ito}(2020{\natexlab{b}})}]{Ito_2020_3}
{Ito}, Y., Mar. 2020{\natexlab{b}}. {Self-similar orbit-averaged Fokker-Planck
  equation for isotropic spherical dense clusters (iii) Application to Galactic
  globular clusters}. arXiv e-prints, arXiv:2004.00747.

\bibitem[{Ito et~al.(2018)Ito, Poje, and Lancellotti}]{Ito_2018}
Ito, Y., Poje, A., Lancellotti, C., jan 2018. Very-large-scale spectral
  solutions for spherical polytropes of index m > 5 and the isothermal sphere.
  New Astronomy 58, 15--28.
\newline\urlprefix\url{https://doi.org/10.1016%2Fj.newast.2017.07.003}

\bibitem[{Josephson(1967)}]{Josephson_1967}
Josephson, B.~D., oct 1967. Inequality for the specific heat: I. derivation.
  Proceedings of the Physical Society 92~(2), 269--275.
\newline\urlprefix\url{https://doi.org/10.1088%2F0370-1328%2F92%2F2%2F301}

\bibitem[{Kang and He(2011)}]{Kang_2011}
Kang, D.-B., He, P., jul 2011. Fluid-like entropy and equilibrium statistical
  mechanics of self-gravitating systems. Monthly Notices of the Royal
  Astronomical Society, no--no.
\newline\urlprefix\url{https://doi.org/10.1111%2Fj.1365-2966.2011.18920.x}

\bibitem[{Katz(1978)}]{Katz_1978}
Katz, J., aug 1978. On the number of unstable modes of an equilibrium. Monthly
  Notices of the Royal Astronomical Society 183~(4), 765--770.
\newline\urlprefix\url{https://doi.org/10.1093%2Fmnras%2F183.4.765}

\bibitem[{Katz(1979)}]{Katz_1979}
Katz, J., dec 1979. On the number of unstable modes of an equilibrium - {II}.
  Monthly Notices of the Royal Astronomical Society 189~(4), 817--822.
\newline\urlprefix\url{https://doi.org/10.1093%2Fmnras%2F189.4.817}

\bibitem[{Katz(1980)}]{Katz_1980}
Katz, J., mar 1980. Stability limits for `isothermal' cores in globular
  clusters. Monthly Notices of the Royal Astronomical Society 190~(3),
  497--507.
\newline\urlprefix\url{https://doi.org/10.1093%2Fmnras%2F190.3.497}

\bibitem[{{Katz} and {Taff}(1983)}]{Katz_1983}
{Katz}, J., {Taff}, L.~G., Jan 1983. {Stability limits for 'isothermal' cores
  in globular cluster models - Two-component systems}. Apj 264, 476--484.

\bibitem[{King(1966)}]{King_1966}
King, I.~R., feb 1966. The structure of star clusters. {III}. some simple
  dvriamical models. The Astronomical Journal 71, 64.
\newline\urlprefix\url{https://doi.org/10.1086%2F109857}

\bibitem[{Komatsu et~al.(2010)Komatsu, Kiwata, and Kimura}]{Komatsu_2010}
Komatsu, N., Kiwata, T., Kimura, S., aug 2010. Thermodynamic properties of an
  evaporation process in self-{gravitatingN}-body systems. Physical Review E
  82~(2).
\newline\urlprefix\url{https://doi.org/10.1103%2Fphysreve.82.021118}

\bibitem[{Komatsu et~al.(2012)Komatsu, Kiwata, and Kimura}]{Komatsu_2012}
Komatsu, N., Kiwata, T., Kimura, S., feb 2012. Transition of velocity
  distributions in collapsing self-{gravitatingN}-body systems. Physical Review
  E 85~(2).
\newline\urlprefix\url{https://doi.org/10.1103%2Fphysreve.85.021132}

\bibitem[{Louis and Spurzem(1991)}]{Louis_1991}
Louis, P.~D., Spurzem, R., 1991. Anisotropic gaseous models for the evolution
  of star clusters. Monthly Notices of the Royal Astronomical Society 251~(3),
  408--426.
\newline\urlprefix\url{http://dx.doi.org/10.1093/mnras/251.3.408}

\bibitem[{Lynden-Bell(1999)}]{Lynden_Bell_1999}
Lynden-Bell, D., feb 1999. Negative specific heat in astronomy, physics and
  chemistry. Physica A: Statistical Mechanics and its Applications 263~(1-4),
  293--304.
\newline\urlprefix\url{https://doi.org/10.1016%2Fs0378-4371%2898%2900518-4}

\bibitem[{Lynden-Bell and Eggleton(1980)}]{Lynden_Bell_1980}
Lynden-Bell, D., Eggleton, P., jul 1980. On the consequences of the
  gravothermal catastrophe. Monthly Notices of the Royal Astronomical Society
  191~(3), 483--498.
\newline\urlprefix\url{http://dx.doi.org/10.1093/mnras/191.3.483}

\bibitem[{Lynden-Bell and Lynden-Bell(1977)}]{Lynden_Bell_1977}
Lynden-Bell, D., Lynden-Bell, R.~M., dec 1977. On the negative specific heat
  paradox. Monthly Notices of the Royal Astronomical Society 181~(3), 405--419.
\newline\urlprefix\url{https://doi.org/10.1093%2Fmnras%2F181.3.405}

\bibitem[{Lynden-Bell and Lynden-Bell(2008)}]{Lynden_Bell_2008}
Lynden-Bell, D., Lynden-Bell, R.~M., may 2008. Negative heat capacities do
  occur. comment on {\textquotedblleft}critical analysis of negative heat
  capacities in nanoclusters{\textquotedblright} by michaelian k. and
  santamar{\'{\i}}a-holek i. {EPL} (Europhysics Letters) 82~(4), 43001.
\newline\urlprefix\url{https://doi.org/10.1209%2F0295-5075%2F82%2F43001}

\bibitem[{Michaelian and Santamar{\'{\i}}a-Holek(2007)}]{Michaelian_2007}
Michaelian, K., Santamar{\'{\i}}a-Holek, I., jul 2007. Critical analysis of
  negative heat capacity in nanoclusters. Europhysics Letters ({EPL}) 79~(4),
  43001.
\newline\urlprefix\url{https://doi.org/10.1209%2F0295-5075%2F79%2F43001}

\bibitem[{Michaelian and Santamar{\'{\i}}a-Holek(2015)}]{Michaelian_2015}
Michaelian, K., Santamar{\'{\i}}a-Holek, I., oct 2015. Dynamics and
  thermodynamics of nanoclusters. Entropy 17~(12), 7133--7148.
\newline\urlprefix\url{https://doi.org/10.3390%2Fe17107133}

\bibitem[{Michaelian and Santamar{\'{\i}}a-Holek(2017)}]{Michaelian_2017}
Michaelian, K., Santamar{\'{\i}}a-Holek, I., jun 2017. Invalid microstate
  densities for model systems lead to apparent violation of thermodynamic law.
  Entropy 19~(7), 314.
\newline\urlprefix\url{https://doi.org/10.3390%2Fe19070314}

\bibitem[{Michie(1962)}]{Michie_1962}
Michie, R.~W., aug 1962. On the distribution of high energy stars in spherical
  stellar systems. Monthly Notices of the Royal Astronomical Society 125~(2),
  127--139.
\newline\urlprefix\url{https://doi.org/10.1093%2Fmnras%2F125.2.127}

\bibitem[{Padmanabhan(1989)}]{Padmanabhan_1989}
Padmanabhan, T., nov 1989. Antonov instability and gravothermal catastrophe -
  revisited. The Astrophysical Journal Supplement Series 71, 651.
\newline\urlprefix\url{https://doi.org/10.1086%2F191391}

\bibitem[{Posch et~al.(1990)Posch, Narnhofer, and Thirring}]{Posch_1990}
Posch, H.~A., Narnhofer, H., Thirring, W., aug 1990. Dynamics of unstable
  systems. Physical Review A 42~(4), 1880--1890.
\newline\urlprefix\url{https://doi.org/10.1103%2Fphysreva.42.1880}

\bibitem[{Posch and Thirring(2005)}]{Posch_2005}
Posch, H.~A., Thirring, W., dec 2005. Stellar stability by thermodynamic
  instability. Physical Review Letters 95~(25).
\newline\urlprefix\url{https://doi.org/10.1103%2Fphysrevlett.95.251101}

\bibitem[{Prabhu and Sharp(2006)}]{Prabhu_2006}
Prabhu, N.~V., Sharp, K.~A., may 2006. Heat capacity in proteins. {ChemInform}
  37~(19).
\newline\urlprefix\url{https://doi.org/10.1002%2Fchin.200619262}

\bibitem[{Schmidt et~al.(2001)Schmidt, Kusche, Hippler, Donges, Kronmüller,
  von Issendorff, and Haberland}]{Schmidt_2001}
Schmidt, M., Kusche, R., Hippler, T., Donges, J., Kronmüller, W., von
  Issendorff, B., Haberland, H., feb 2001. Negative heat capacity for a cluster
  of 147 sodium atoms. Physical Review Letters 86~(7), 1191--1194.
\newline\urlprefix\url{https://doi.org/10.1103%2Fphysrevlett.86.1191}

\bibitem[{Spitzer and Shapiro(1972)}]{Spitzer_1972}
Spitzer, L.~J., Shapiro, S.~L., may 1972. Random gravitational encounters and
  the evolution of spherical systems. {III}. halo. The Astrophysical Journal
  173, 529.
\newline\urlprefix\url{https://doi.org/10.1086%2F151442}

\bibitem[{Spitzer(1988)}]{Spitzer_1988}
Spitzer, L.~S., jan 1988. Dynamical Evolution of Globular Clusters. Walter de
  Gruyter {GmbH}.
\newline\urlprefix\url{http://dx.doi.org/10.1515/9781400858736}

\bibitem[{Srivastava(2001)}]{Srivastava_2001}
Srivastava, B., aug 2001. Multifragmentation and the phase transition: A
  systematic study of the multifragmentation of 1a {GeV} au, la and kr. Pramana
  57~(2-3), 301--313.
\newline\urlprefix\url{https://doi.org/10.1007%2Fs12043-001-0040-x}

\bibitem[{Takahashi(1993)}]{Takahashi_1993}
Takahashi, K., 1993. Self-similar solutions of the orbit-averaged fokker-planck
  equation: Application of the generalized variational principle. Publications
  of the Astronomical Society of Japan 45, 789--793.

\bibitem[{Taruya and Sakagami(2002)}]{Taruya_2002}
Taruya, A., Sakagami, M., apr 2002. Gravothermal catastrophe and {Tsallis}'
  generalized entropy of self-gravitating systems. Physica A: Statistical
  Mechanics and its Applications 307~(1-2), 185--206.
\newline\urlprefix\url{http://dx.doi.org/10.1016/s0378-4371(01)00622-7}

\bibitem[{Thirring(1970)}]{Thirring_1970}
Thirring, W., aug 1970. Systems with negative specific heat. Zeitschrift für
  Physik A Hadrons and nuclei 235~(4), 339--352.
\newline\urlprefix\url{https://doi.org/10.1007%2Fbf01403177}

\bibitem[{Thirring et~al.(2003)Thirring, Narnhofer, and Posch}]{Thirring_2003}
Thirring, W., Narnhofer, H., Posch, H.~A., sep 2003. Negative specific heat,
  the thermodynamic limit, and ergodicity. Physical Review Letters 91~(13).
\newline\urlprefix\url{https://doi.org/10.1103%2Fphysrevlett.91.130601}

\bibitem[{Tsallis(2009)}]{Tsallis_2009}
Tsallis, C., 2009. Introduction to Nonextensive Statistical Mechanics. Springer
  Science $\mathplus$ Business Media.
\newline\urlprefix\url{http://dx.doi.org/10.1007/978-0-387-85359-8}

\end{thebibliography}


\end{document}